\documentclass[11pt]{article}
\usepackage{verbatim,color,amssymb,epsfig,epsf,graphicx,indentfirst,caption,lscape,multirow}
\usepackage{amscd,amsmath,amstext,amsfonts,amsbsy,amsthm, rotating, booktabs}
\RequirePackage{url}
\newcommand{\ud}{\mathrm{d}}
\newcommand{\logit}{\mathrm{logit}}

\newcommand{\V}{\mathbb{V}\mathrm{ar}}
\newcommand{\COV}{\mathbb{C}\mathrm{ov}}

\newcommand{\Pro}{\mathrm{Pr}}

\def\Real{I\!\!R}

\begin{document}
\begin{center}
{\Large{\bf A pseudo-likelihood approach for multivariate meta-analysis of test accuracy studies with multiple thresholds}}\\
{Annamaria Guolo* and Duc Khanh To}
\\{University of Padova\\
Via Cesare Battisti 241/243, I-35121, Padova, Italy}
\\{* Corresponding author: annamaria.guolo@unipd.it}
\end{center}

\begin{center}
SUMMARY 
\end{center}
Multivariate meta-analysis of test accuracy studies when tests are evaluated in terms of sensitivity and specificity at more than one threshold represents an effective way to synthesize results by fully exploiting the data, if compared to univariate meta-analyses performed at each threshold independently. The approximation of logit transformations of sensitivities and specificities at different thresholds through a normal multivariate random-effects model is a recent proposal, that straightforwardly extends the bivariate models well recommended for the one threshold case. However, drawbacks of the approach, such as poor estimation of the within-study correlations between sensitivities and between specificities and severe computational issues, can make it unappealing. We propose an alternative method for inference on common diagnostic measures using a pseudo-likelihood constructed under a working independence assumption between sensitivities and between specificities at different thresholds in the same study. The method does not require within-study correlations, overcomes the convergence issues and can be effortlessly implemented. Simulation studies highlight a satisfactory performance of the method, remarkably improving the results from the multivariate normal counterpart under different scenarios. The pseudo-likelihood approach is illustrated in the evaluation of a test used for diagnosis of pre-eclampsia as a cause of maternal and perinatal morbidity and mortality.
\\[2ex]

\section{Introduction}\label{sec:intro}
Meta-analysis of test accuracy studies represents a widely accepted approach in medical investigations to evaluate the performance of a diagnostic test compared to a reference standard. 

The majority of the studies evaluate a test in terms of sensitivity and specificity, that is, in terms of its capability to distinguish diseased and nondiseased subject with respect to a single threshold. Within this framework, bivariate random effects models are a well established instrument \cite{arends2008bivariate, hamza2008meta, chen2017composite}.
In a more realistic scenario, test performance is evaluated at multiple thresholds or, equivalently, at ordered categories, representing, for example, increasing levels of the severity of a pathology. Correspondingly, studies report the classification of diseased and nondiseased patients provided by the test at each threshold. A straightforward, although questionable, approach dichotomizes the test results into two categories and proceeds with a standard bivariate meta-analysis. More efficient solutions exploit the induced correlation between sensitivities and between specificities at different thresholds within a study. 
Recently, Riley et al. \cite{riley2014meta} proposed an extension of the bivariate random-effects approach to handle studies with multiple and possibly missing thresholds. The resulting multivariate normal accounts for the possibility of different thresholds per study and guarantees the monotonicity of sensitivity and specificity for increasing thresholds. In this way, the proposal overcomes drawbacks in previous studies, e.g., Hamza et al. \cite{hamza2009multivariate} and Putter et al. \cite{putter2010meta}. Despite the advantages, the approach in Riley et al. \cite{riley2014meta} suffers from several drawbacks. The within-study covariance components are typically not available and they need to be estimated at a preliminary step. In addition, nonconvergence often occurs and estimates of the covariance parameters can be close to the boundary of the parameter space. 

This paper proposes a novel pseudo-likelihood approach inspired by the composite likelihood methodology \cite{varin2011}, in this way avoiding the limitations affecting the model in Riley et al. \cite{riley2014meta} while retaining many of its advantages. The proposed model is developed under a working independence assumption between sensitivities and between specificities at different thresholds in the same study, with no need to provide preliminary estimates of the within-study covariances as required instead in Riley et al. \cite{riley2014meta}. The methodology can be implemented with no computational effort and is not prone to convergence problems. Valid inference for the summary measures of the accuracy of diagnostic tests are obtained through a proper evaluation of the variability of the resulting estimators. The performance of the pseudo-likelihood approach for inference on the summary specificity and sensitivity, the summary receiver operating characteristic (SROC) curve and the area underneath is compared to that of the two-steps likelihood approach in Riley et al. \cite{riley2014meta} through an extensive simulation study covering different scenarios, with varying sample size and variance components of the random-effects. The pseudo-likelihood approach is finally applied to a meta-analysis of diagnostic test in patients with suspected pre-eclampsia, a condition related to maternal and perinatal morbidity and mortality.

\section{Background}\label{sec:bgs}

Consider a meta-analysis of $K$ test accuracy studies, each of them providing the measure of a continuous test $T$ result on $n_{1k}$ diseased patients and $n_{0k}$ nondiseased patients, with $n_{1k} + n_{0k}=n_k$, $k=1, \ldots, K$. Let $x_k$ be a threshold value in study $k$ such that a patient is classified as positive if his/her measured test value $t_{ki}$, $i=1,\ldots, n_k$, is equal or larger than $x_k$, $t_{ki} \geq x_k$, or negative otherwise, $t_{ki} < x_k$. For generality purposes, it can be assumed that the number of thresholds varies across the studies, with $m_k$ denoting the number of thresholds $\{x_{k1}, \cdots, x_{km_k}\}$ in study $k$ and $x_{kj} < x_{{kj} + 1}, \ j = 1, \ldots, m_k$. Suppose that the true disease status of each patient is known as provided by a reference standard. A diagnostic test is evaluated in terms of sensitivity, that is, the conditional probability of testing positive in diseased subjects, and specificity, that is, the conditional probability of testing negative in nondiseased subjects. With reference to study $k$, the sensitivity at threshold $x_{kj}$ is $Se\left(x_{kj}\right) = Se_{kj}=\Pr \left(T_k \geq x_{kj} | \, \text{diseased}\right)$ and the specificity is $Sp\left(x_{kj}\right) = Sp_{kj}=\Pr \left(T_k < x_{kj} | \, \text{nondiseased}\right)$. Empirical estimates of $Se_{kj}$ and $Sp_{kj}$ are provided by the percentage of true positives $TP_{kj}$ and true negatives $TN_{kj}$ at threshold $x_{kj}$ in study $k$, namely,
\[
\widehat{Se}_{kj} = \frac{1}{n_{1k}} \sum_{i = 1}^{n_{1k}} \mathrm{I}\left(T_{ki} \ge x_{kj}\right) = \frac{TP_{kj}}{n_{1k}}, \, \widehat{Sp}_{kj} = \frac{1}{n_{0k}} \sum_{i = 1}^{n_{0k}} \mathrm{I}\left(T_{ki} < x_{kj}\right) = \frac{TN_{kj}}{n_{0k}},
\]
where $\mathrm{I}(\cdot)$ is the indicator function. Typically, the accuracy of a test is expressed in terms of logit transformations $Y_{1kj} = \logit\left({Se}_{kj}\right)$ and $Y_{0kj}= \logit\left(Sp_{kj}\right)$. Hereafter, the estimator of $Y_{1kj}$ and $Y_{0kj}$ will be denoted by $\hat Y_{1kj}$ and $\hat Y_{0kj}$, respectively.

\subsection{Single threshold}\label{sec:bgs:1}
In case of single threshold, with $Y_{1kj}= Y_{1k}$ and $Y_{0kj}=Y_{0k}$, the bivariate mixed-effects model following Reitsma et al. \cite{reitsma2005bivariate} and Arends et al. \cite{arends2008bivariate} is a well-established approach for inference on the overall sensitivity and specificity and on the SROC curve. The model distinguishes a between-study and a within-study structure, as follows. The between-study model considers a joint normal distribution for $\left(Y_{1k}, -Y_{0k}\right)^\top$, with $-Y_{0k}=\logit\left(1-Sp_{k}\right)$,
\begin{equation}\label{eqn:bi_ma_bet}
\begin{pmatrix}
Y_{1k}\\
-Y_{0k}
\end{pmatrix} \sim \mathcal{N}_2 \left( \begin{pmatrix}
\mu_1 \\
\mu_0
\end{pmatrix}, \begin{pmatrix}
\sigma^2_{1} & \lambda \sigma_1 \sigma_0 \\
\lambda \sigma_1 \sigma_0  & \sigma^2_0
\end{pmatrix}
\right),
\end{equation}
where $\mu_1$ and $\mu_0$ are the means over the studies, $\sigma^2_{1}$ and $\sigma^2_{0}$ are the between-study variances and $\lambda$ is the correlation coefficient. Typically, the within-study model relating the observed measures and the true measures of test accuracy is approximated by a normal distribution 
\begin{equation*}
\begin{pmatrix}
\hat{Y}_{1k} \\
-\hat{Y}_{0k}
\end{pmatrix} \sim \mathcal{N}_2 \left( \begin{pmatrix}
Y_{1k}\\
-Y_{0k}
\end{pmatrix}, \left(
\begin{array}{c c}
n_{1k}^{-1}\widehat{Se}_k^{-1}(1 - \widehat{Se}_k)^{-1} & 0 \\ [8pt]
0 & n_{0k}^{-1}\widehat{Sp}_k^{-1}(1 - \widehat{Sp}_k)^{-1}
\end{array}\right)
\right),
\end{equation*}
where the covariance matrix is diagonal as sensitivity and the specificity are estimated on different patients. Non-zero entries are estimated from each study. Marginally,
\begin{equation*}
\label{eqn:bi_ma_marg}
\begin{pmatrix}
\hat{Y}_{1k} \\
-\hat{Y}_{0k}
\end{pmatrix} \sim  \mathcal{N}_2 \left( \begin{pmatrix}
\mu_1 \\
\mu_0
\end{pmatrix}, \left(
\begin{array}{cc}
\sigma^2_{1} + n_{1k}^{-1}\widehat{Se}_k^{-1}(1 - \widehat{Se}_k)^{-1} & \lambda \sigma_1 \sigma_0 \\
\lambda \sigma_1 \sigma_0  & \sigma^2_0 + n_{0k}^{-1}\widehat{Sp}_k^{-1}(1 - \widehat{Sp}_k)^{-1}
\end{array}
\right)
\right).
\end{equation*}
Likelihood inference on $\left(\mu_1, \mu_0, \sigma^2_1, \sigma^2_0, \lambda\right)^\top$ takes advantage from the likelihood function being in closed-form.

The literature (e.g., \cite{hamza2008binomial}) suggests the alternative exact within-study model by specifying the observed true positives and false positives as results of binomial variables. This gives rise to a likelihood function not in closed-form. Computational issues arise with respect to the approximate specification, namely, risk of non-positive definite covariance matrix and estimates of the variance or correlation parameters on the boundary of the parameter space (\cite{chen2017composite,hamza2008meta,takwoingi2017performance}), especially for small $K$.
\subsection{Multiple thresholds}
\label{sec:bgs:2}
In case of multiple thresholds, Riley et al. \cite{riley2014meta} propose a multivariate normal random-intercept model as an extension of the bivariate solution in Arends et al. \cite{arends2008bivariate}. Let $s^2_{1kj} = \V(Y_{1kj}), s^2_{0kj} = \V(Y_{0kj}), s_{1kjj'} = \COV(Y_{1kj}, Y_{1kj'})$ and $s_{0kjj'} = \COV(Y_{0kj}, Y_{0kj'})$ be the variances and covariances for $Y_{1kj}$ and $Y_{0kj}$ at threshold $x_{kj}$ for study $k$, respectively. Then, (logit) sensitivities and specificities for study $k$ at different thresholds follow a normal distribution with a mean specified as linear function of the thresholds, 
\begin{equation}
\begin{pmatrix}
Y_{1k1} \\ \vdots \\ Y_{1km_k} \\ Y_{0k1} \\ \vdots \\ Y_{0km_k} 
\end{pmatrix} \sim \mathcal{N}_{2m_k} \begin{pmatrix} \begin{pmatrix}
\alpha_{1k} + \gamma_1 x_{k1} \\ \vdots \\ \alpha_{1k} + \gamma_1 x_{km_k} \\ \alpha_{0k} + \gamma_0 x_{k1} \\ \vdots \\ \alpha_{0k} + \gamma_0 x_{km_k} 
\end{pmatrix}, \Omega_K
\end{pmatrix},
\label{md1}
\end{equation}
with 
$$
\Omega_K = \begin{pmatrix}
s^2_{1k1} & \ldots & s_{1k1m_k} & 0 & \ldots & 0  \\
\vdots & \ddots & \vdots  & \vdots & \ddots & \vdots \\
s_{1km_k1} & \ldots & s^2_{1km_k} & 0 & \ldots & 0 \\
0 & \ldots & 0 & s^2_{0k1} & \ldots & s_{0k1m_k} \\
\vdots & \ddots & \vdots & \vdots & \ddots & \vdots \\
0 & \ldots & 0 & s_{0k m_k 1} & \ldots & s^2_{0km_k} 
\end{pmatrix}.
$$
Random intercepts follow a normal distribution, 
\begin{eqnarray}
\begin{pmatrix}
\alpha_{1k} \\ \alpha_{0k}
\end{pmatrix} &\sim& \mathcal{N}_2 \begin{pmatrix} \begin{pmatrix}
\alpha_1 \\ \alpha_0
\end{pmatrix}, \begin{pmatrix}
\tau_1^2  & \rho \tau_1 \tau_0 \\
\rho \tau_1 \tau_0 & \tau_0^2
\end{pmatrix} \end{pmatrix},
\label{md2}
\end{eqnarray}
where parameters $\alpha_0$ and $\alpha_1$ are average intercepts and $\tau_0^2, \tau_1^2$  and $\rho$ represent between-study variances and between-study correlation, respectively.
Constraints $\gamma_0 \ge 0$ and $\gamma_1 \le 0$ guarantee monotonicity of sensitivities and specificities at increasing thresholds. In addition, the inclusion of studies with different thresholds is allowed. In this way, the proposal in Riley et al. \cite{riley2014meta} overcomes two drawbacks affecting the extension of the bivariate model suggested in Hamza et al. \cite{hamza2009multivariate} and characterized by an (exact) multinomial within-study specification. 

Riley et al. \cite{riley2014meta} use a two-step approach to perform inference on the parameter vector $\boldsymbol{\theta}=\left(\alpha_1, \alpha_0, \gamma_1, \gamma_0, \tau^2_1, \tau^2_0, \rho\right)^\top$. In the first step, the within-study covariance matrix components for each study are estimated by maximizing the multinomial likelihood function of the probabilities that test values fall between specific thresholds. In the second step, inference relies on maximum likelihood (ML) or restricted maximum likelihood (REML), after substituting the within study covariance matrix components with the estimates from the first step. The problem of missing thresholds across studies is faced by assuming a missing at random mechanism and adopting a data-augmentation approach. A continuity correction is applied for sensitivity or specificity equal to zero.

Despite some advantages, the approach in Riley et al. \cite{riley2014meta} suffers from several drawbacks that can substantially limit its application. The estimation of the within-study covariances is a necessary step as the information is not typically available from the single studies included in a meta-analysis. However, the variability associated to the covariances estimation should be taken into account, otherwise inferential conclusions on $\boldsymbol{\theta}$ can be misleading. From a computational point of view, although the approach is less cumbersome than that in Hamza et al. \cite{hamza2009multivariate}, as the likelihood function for model specification (\ref{md1})--(\ref{md2}) has a closed-form expression, nonconvergence problems are substantial, especially when the number of thresholds per study increases. They are mainly related to the estimation of variance/covariance components, typically resulting on the boundary of the parameter space.
\section{Pseudo-likelihood approach}
\label{sec:proposal_pseudo}
We suggest to perform inference in meta-analysis of test accuracy studies in case of multiple and possibly missing thresholds through a pseudo-likelihood approach which takes advantage of the methodology developed in Riley et al. \cite{riley2014meta}, while reducing its computational issues and overcoming limits of a two-step procedure. To this aim, we propose a pseudo-likelihood approach starting from models (\ref{md1})--(\ref{md2}) under a working independence assumption between sensitivities and between specificities at different thresholds within each study included in the meta-analysis, in the spirit of the composite likelihood approach in Varin et al. \cite{varin2011}. See also \cite{chen2015alternative} and \cite{chen2017composite}. According to the working independence assumption, the within-study covariance matrix in (\ref{md1}) can be re-written as a diagonal matrix
$$
\Omega_K^{pseudo} = diag\left\{
s^2_{1k1}, \ldots,  s^2_{1km_k}, \ldots, s^2_{0k1}, \ldots, s^2_{0km_k}
\right\}.$$ 
The non-zero entries are considered as nuisance parameters and they can be substituted by their sample-based counterpart  
$$s^2_{1kj} = n_{1k}^{-1}\widehat{Se}_{kj}^{-1}(1 - \widehat{Se}_{kj})^{-1}, \qquad s^2_{0kj} = n_{0k}^{-1} \widehat{Sp}_{kj}^{-1}(1 - \widehat{Sp}_{kj})^{-1}.
$$
The approach has connections with the pseudo-likelihood approach in Gong and Samaniego \cite{gong1981}, although we consider the within-study covariance matrix components only as nuisance parameters, while the between-study counterpart remains subject to the inferential interest. 

According to model specification in Section~\ref{sec:bgs:2}, the marginal distribution of $\mathbf{Y}_k$ follows a multivariate $2m_k$ normal, with covariance matrix
\[
\Sigma_k = \mathbb{I}_{k} \begin{pmatrix}
\tau_1^2  & \rho \tau_1 \tau_0 \\
\rho \tau_1 \tau_0 & \tau_0^2
\end{pmatrix} \mathbb{I}^\top_{k} + \Omega_k,
\]
where $\mathbb{I}_{k} = \begin{pmatrix}
\boldsymbol{1}_{m_k \times 1} & \boldsymbol{0}_{m_k \times 1} \\
\boldsymbol{0}_{m_k \times 1} & \boldsymbol{1}_{m_k \times 1} 
\end{pmatrix}$, $\boldsymbol{1}_{m_k \times 1}$ is a $m_k$-dimensional vector of ones and $\boldsymbol{0}_{m_k \times 1}$ is a $m_k$-dimensional vector of zeros. Inference on the parameter vector $\boldsymbol{\theta}$ can be performed through maximum likelihood. Let $\boldsymbol{X}_k = \left(x_{k1}, \cdots, x_{km_k}\right)^\top$, let
\[
\mathbb{Z}_{k} = \begin{pmatrix}
\boldsymbol{1}_{m_k \times 1} & \boldsymbol{0}_{m_k \times 1} & \boldsymbol{X}_k & \boldsymbol{0}_{m_k \times 1} \\
\boldsymbol{0}_{m_k \times 1} & \boldsymbol{1}_{m_k \times 1} & \boldsymbol{0}_{m_k \times 1} & \boldsymbol{X}_k
\end{pmatrix},
\]
and let $\boldsymbol{\beta} = \left(\alpha_1, \alpha_0, \gamma_1, \gamma_0\right)^\top$. The log-likelihood contribution of study $k$ is 
\begin{equation}
\ell_k(\boldsymbol{\theta}) = -\frac{1}{2} \log |\Sigma_k| - \frac{1}{2}\left(\mathbf{Y}_k - \mathbb{Z}_{k} \boldsymbol{\beta}\right)^\top \Sigma_k^{-1} \left(\mathbf{Y}_k - \mathbb{Z}_{k} \boldsymbol{\beta}\right).
\label{log-lik}
\end{equation}
Alternatively, inference can be based on REML, with the restricted log-likelihood contribution for the covariance components given by (\cite{viechtbauer2005}, Chapter 7 in \cite{schwarzer2015})
\begin{equation}
\ell_k(\boldsymbol{\theta}) - \frac{1}{2K} \log \left|\sum_{k = 1}^{K}\mathbb{Z}_{k}^\top \Sigma_k^{-1} \mathbb{Z}_{k}  \right|.
\label{restr-log-lik}
\end{equation}

Properties of the maximum pseudo-likelihood estimator $\boldsymbol{\widehat{\theta}}$ derive from the results on composite likelihood, see, for example, \cite{varin2011}. The maximum pseudo-likelihood estimator asymptotically follows a normal distribution with mean $\boldsymbol \theta$ and covariance matrix $\COV(\boldsymbol{\widehat{\theta}})$
$$
\boldsymbol{\widehat{\theta}} \sim N(\boldsymbol \theta, \COV(\boldsymbol{\widehat{\theta}})),
$$
where $\COV(\boldsymbol{\widehat{\theta}})$ can be estimated using the sandwich formula. In this way, both the variability associated to the working independence assumption and the uncertainty related to the estimation of the within-study variances in $\Omega_K^{pseudo}$ can be taken into account. The sandwich estimator of the covariance matrix (\cite{liang1986longitudinal,kauermann2001note,mancl2001covariance}) of $\boldsymbol{\widehat{\theta}}$ is obtained as
\begin{equation}
\widehat\COV(\boldsymbol{\widehat{\theta}})= \left.K^{-1} J^{-1}(\boldsymbol{\theta})I(\boldsymbol{\theta})J^{-1} (\boldsymbol{\theta})\right|_{\boldsymbol{\theta}= \boldsymbol{\widehat{\theta}}},
\label{var:sand_type}
\end{equation}
where
$$
J(\boldsymbol{\theta}) = K^{-1} E \left\{ \frac{\partial^2 }{\partial \boldsymbol{\theta} \partial \boldsymbol{\theta}^\top} \ell_k(\boldsymbol{\theta}) \right\}, \ \
I(\boldsymbol{\theta}) = K^{-1} E \left[\frac{\partial}{\partial \boldsymbol{\theta} } \ell_k(\boldsymbol{\theta})\left\{\frac{\partial}{\partial \boldsymbol{\theta} } \ell_k(\boldsymbol{\theta}) \right\}^\top \right],
$$
 $\partial f(x) /\partial x$ is the derivative of $f(x)$ with respect to $x$ and $E\{X\}$ is the expectation of $X$. Matrices $I(\boldsymbol{\theta})$ and $J(\boldsymbol{\theta})$ can be consistently estimated by their empirical counterpart.

Measures of the accuracy of the diagnostic test (e.g., \cite{arends2008bivariate}) can be derived starting from $\boldsymbol{\hat\theta}$. Following Pepe \cite{pep}, the SROC curve can be obtained by substituting the estimates of the parameters in the functional form  
\begin{equation}\label{eqn:sroc}
\mathrm{SROC}\left(t; \alpha_1, \alpha_0, \gamma_1, \gamma_0\right) = \logit^{-1} \left\{\alpha_1 + \gamma_1 \frac{\logit(1 - t) - \alpha_0}{\gamma_0} \right\},
\end{equation}
for $t \in (0,1)$, and its point-wise variance for given $t$ can be obtained through the delta method. See the Supplementary Material for details.
The estimate of the area under SROC curve (AUSC) and the associated variance can be obtained in a similar way. 

When constructing ROC curves, a requirement largely emphasized refers to the ROC curve being concave, or \emph{proper} in the biomedical terminology, see the discussion in Gneiting and Vogel \cite{gneiting}. ROC curves within meta-analysis context are typically concave in shape, as Walter \cite{walter2002} and Pepe \cite{pep} point out. The proposed SROC curve in (\ref{eqn:sroc}) is guaranteed to be proper when $\gamma_1=-\gamma_0$. Far from that case, there is the possibility of non-concavity in bottom-left corner or in the top-right corner of the SROC space, in this way meeting the results in Walter \cite{walter2002}. See the Supplementary Material for details.

Alternatives specifications of the SROC curves are possible, see the suggestions in Arends et al. \cite{arends2008bivariate}, although referred to the single threshold case. A common specification is the one proposed in Moses et al. \cite{moses1993combining}, based on a regression model with dependent and independent variables given by the difference and the sum of the logit sensitivity and logit false positive rate, respectively. The properties of the resulting SROC curve in terms of concavity is investigated in Walter \cite{walter2002}. The connections between the SROC specification in Moses et al. \cite{moses1993combining} and our formulation in (\ref{eqn:sroc}), discussion about concavity included, are reported in  the Supplementary Material.

Similarly to the approach in Riley et al. \cite{riley2014meta}, the pseudo-likelihood method allows to perform meta-analysis from studies about accuracy of diagnostic tests evaluated at different thresholds, including studies with a single threshold. Studies with missing thresholds information can be included as well, see the Supplementary Material for details. The monotonicity of the estimated sensitivities and specificities at increasing thresholds is maintained.

\section{Simulation study}\label{sec:sim}
The performance of the pseudo-likelihood approach is evaluated through simulation and compared to that of the approach in Riley et al. \cite{riley2014meta}. Data simulation follows a two-step procedure. 
Values of the random intercepts $\alpha_{0k}$ and $\alpha_{1k}$ are generated from the bivariate normal distribution in (\ref{md1}), for given $\alpha_0$, $\alpha_1$, $\gamma_0$, $\gamma_1$, $\tau_0^2$, $\tau_1^2$, $\rho$. Then,  for given $K$, true positives and true negatives at threshold $x_j$ are generated from the binomial variables	\[
	TN_{kj} \sim \mathcal{B}\left\{n_{0k}, \logit^{-1}\left(\alpha_{0k} + \gamma_0 x_j\right)\right\}, \, TP_{kj} \sim \mathcal{B}\left\{n_{1k}, \logit^{-1}\left(\alpha_{1k} + \gamma_1 x_j\right)\right\}.
	\]
Within-study numbers of diseased and nondiseased subjects are generated from independent uniform variables on $[10, 500]$. We set $\alpha_0 = 1, \alpha_1 = 2, \gamma_0 = 1.5$ and $\gamma_1 = -2$, in order to guarantee the true AUSC value equal to 0.875. The number of studies $K$ varies in $\{10, 20, 50\}$ in order to consider meta-analysis of small number, moderate number or large number of studies. The number of thresholds per study varies from 3 to 20. We compare a full thresholds case for all the studies included in the meta-analysis and a missing thresholds case. The missing completely at random mechanism for the thresholds is considered, with the number of thresholds sampled in each study so to guarantee that the maximum varies from 3 to 20. Increasing heterogeneity is considered with equal between-study variances $\tau_0^2$ and $\tau_1^2$ assuming values in $\{0.1, 0.5, 1\}$. For each value of the between-study variances, the between-study correlation $\rho$ takes values in $\{0.3, 0.6, 0.9\}$. Five thousand datasets are generated for each combination of between-study variances and correlation.

For each scenario, meta-analysis is performed using the pseudo-likelihood approach with ML estimation (Pseudo-ML) and with REML estimation (Pseudo-REML), the Riley et al. \cite{riley2014meta} approach with ML estimation (Riley-ML) and with REML estimation (Riley-REML). Methods are compared in terms of Monte Carlo means (MCM), Monte Carlo standard deviations (MCSD), average of standard errors (ASE) of the estimators of the parameters in $\boldsymbol{\theta}$ and of AUC and empirical coverages for Wald-type confidence intervals at nominal level 0.95 for parameters $\alpha_0, \alpha_1, \gamma_0, \gamma_1$ and for AUSC. Standard errors of the estimators are obtained using sandwich-type estimator. The alternative bias-corrected sandwich estimator in Kauermann and Carroll \cite{kauermann2001note} suggested in case of small sample size provides similar conclusions and the corresponding results are not reported. 

\begin{figure}
	\begin{center}
			\includegraphics[width=4in]{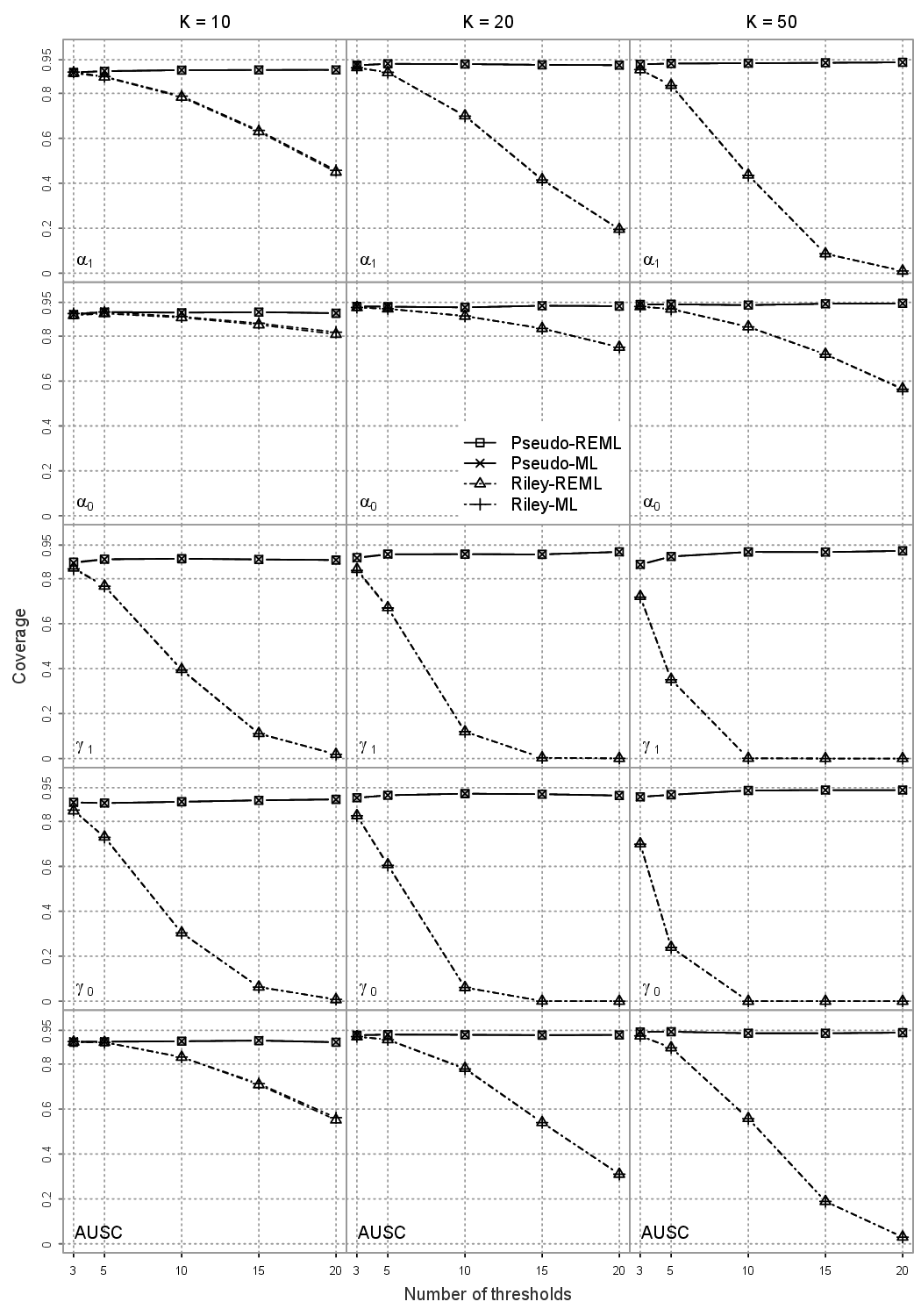}
	\end{center}
	\caption{Empirical coverages of confidence intervals at nominal level 0.95 for $\alpha_0, \alpha_1, \gamma_0, \gamma_1$ and AUSC using Pseudo-REML, Pseudo-ML, Riley-REML and Riley-ML approaches, for $\tau^2_{1} = \tau^2_{0} = 0.1, \rho = 0.6$ and increasing $K$. All thresholds of interest are reported in the studies.}
	\label{fig:1}
\end{figure}

\begin{figure}
	\begin{center}
		\includegraphics[width=4in]{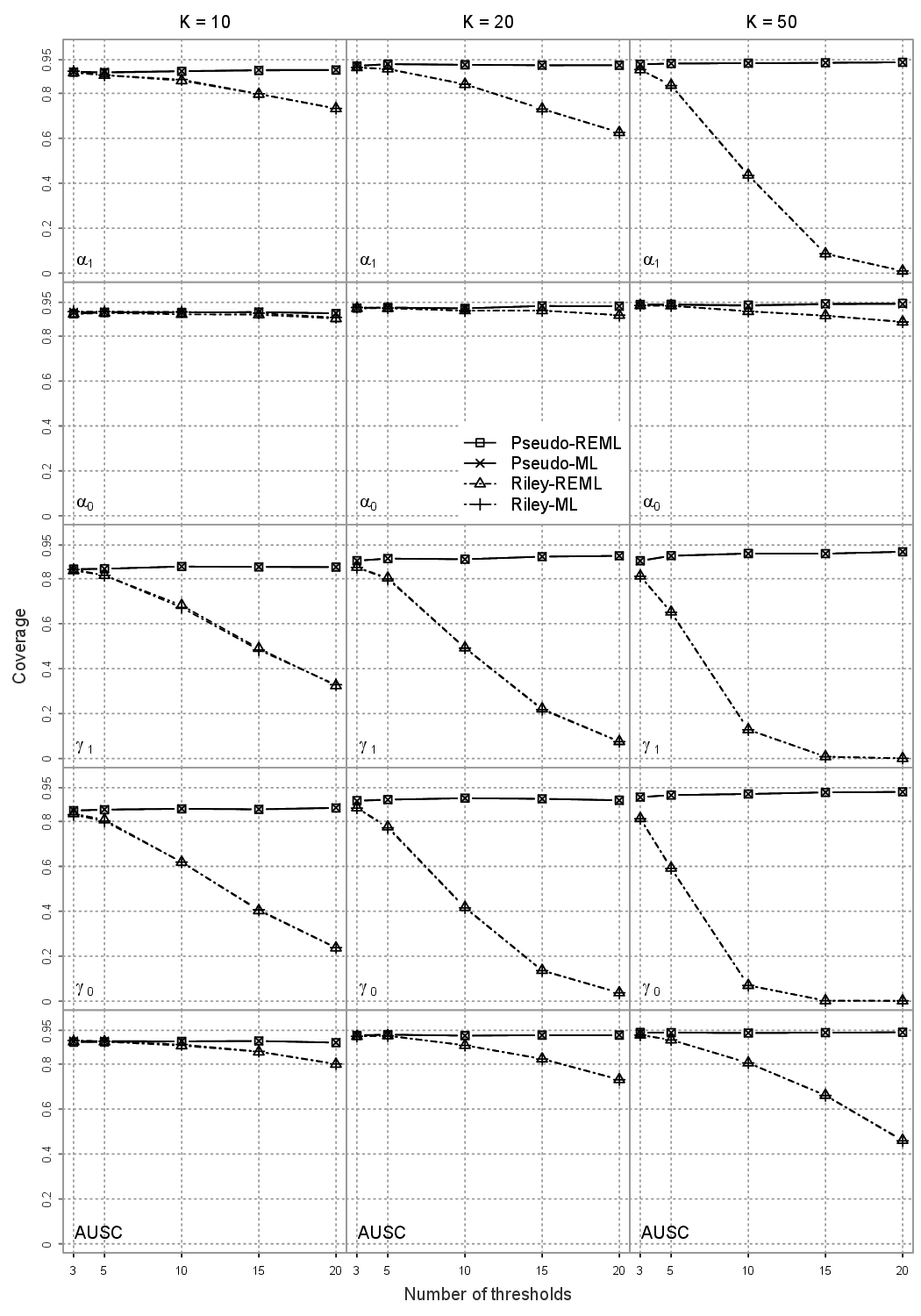}
	\end{center}
	\caption{Empirical coverages of confidence intervals at nominal level 0.95 for $\alpha_0, \alpha_1, \gamma_0, \gamma_1$ and AUSC using Pseudo-REML, Pseudo-ML, Riley-REML and Riley-ML approaches, for $\tau^2_{1} = \tau^2_{0} = 0.1, \rho = 0.6$ and increasing $K$. Different thresholds of interest are reported in the studies.}
	\label{fig:2}
\end{figure}

Simulation results when $\tau^2_{1} = \tau^2_{0} = 0.1$ and $\rho = 0.6$ are reported in Tables \ref{tab:1}--\ref{tab:2} and in Figures \ref{fig:1}--\ref{fig:2}. For space reasons, results in Tables are shown for Pseudo-REML and Riley-REML, as results from the ML counterparts are similar.  
The simulation results for other combinations of the parameters are available in the Supplementary Material.

\begin{sidewaystable}
	\centering
	\caption{Monte Carlo means (MCM), Monte Carlo standard deviations (MCSD), average of standard errors (ASE) for the parameters $\left(\alpha_0, \alpha_1, \gamma_0, \gamma_1, \tau^2_0, \tau^2_1, \rho\right)^\top$ and for AUSC, on the basis of 5,000 replicates of the simulation experiment with $\tau^2_{1} = \tau^2_{0} = 0.1, \rho = 0.6$ and increasing $K$. Maximum number of thresholds equal to 5.} 
	\label{tab:1}
	\begingroup\scriptsize
	\begin{tabular}{lllrrrrrrrrrr}
		\toprule
		& & & & \multicolumn{3}{c}{$K = 10$} & \multicolumn{3}{c}{$K = 20$} & \multicolumn{3}{c}{$K = 50$}  \\
		\cmidrule(l){5-7} \cmidrule(l){8-10}  \cmidrule(l){11-13} 
		& & & TRUE & MCM & MCSD & ASE & MCM & MCSD & ASE & MCM & MCSD & ASE \\
		\midrule
		\multirow{ 16 }{*}{ Full thresholds } & \multirow{ 8 }{*}{ Pseudo-REML } & $\alpha_1$ & 2.000 & 1.986 & 0.116 & 0.107 & 1.987 & 0.080 & 0.079 & 1.986 & 0.052 & 0.051 \\ 
		&  & $\alpha_0$ & 1.000 & 0.996 & 0.111 & 0.103 & 0.995 & 0.076 & 0.076 & 0.993 & 0.049 & 0.049 \\ 
		&  & $\gamma_1$ & $-$2.000 & $-$1.990 & 0.048 & 0.043 & $-$1.988 & 0.033 & 0.032 & $-$1.988 & 0.021 & 0.021 \\ 
		&  & $\gamma_0$ & 1.500 & 1.494 & 0.043 & 0.038 & 1.494 & 0.030 & 0.028 & 1.493 & 0.019 & 0.018 \\ 
		&  & $\tau^2_{1}$ & 0.100 & 0.112 & 0.058 & 0.046 & 0.111 & 0.039 & 0.035 & 0.110 & 0.024 & 0.023 \\ 
		&  & $\tau^2_{0}$ & 0.100 & 0.114 & 0.058 & 0.047 & 0.114 & 0.041 & 0.036 & 0.113 & 0.025 & 0.024 \\ 
		&  & $\rho$ & 0.600 & 0.514 & 0.297 & 0.216 & 0.521 & 0.194 & 0.166 & 0.522 & 0.122 & 0.111 \\ 
		&  & $\mathrm{AUSC}$ & 0.875 & 0.872 & 0.015 & 0.014 & 0.873 & 0.010 & 0.010 & 0.873 & 0.007 & 0.007 \\ 
		\cmidrule(l){2-13}
		& \multirow{ 8 }{*}{ Riley-REML } & $\alpha_1$ & 2.000 & 1.952 & 0.114 & 0.107 & 1.954 & 0.078 & 0.077 & 1.954 & 0.051 & 0.050 \\ 
		&  & $\alpha_0$ & 1.000 & 0.980 & 0.108 & 0.102 & 0.980 & 0.074 & 0.073 & 0.978 & 0.048 & 0.047 \\ 
		&  & $\gamma_1$ & $-$2.000 & $-$1.949 & 0.052 & 0.046 & $-$1.949 & 0.035 & 0.034 & $-$1.949 & 0.022 & 0.022 \\ 
		&  & $\gamma_0$ & 1.500 & 1.448 & 0.047 & 0.042 & 1.448 & 0.032 & 0.031 & 1.448 & 0.020 & 0.020 \\ 
		&  & $\tau^2_{1}$ & 0.100 & 0.096 & 0.054 & 0.045 & 0.096 & 0.037 & 0.033 & 0.095 & 0.023 & 0.022 \\ 
		&  & $\tau^2_{0}$ & 0.100 & 0.095 & 0.054 & 0.046 & 0.096 & 0.038 & 0.034 & 0.095 & 0.023 & 0.022 \\ 
		&  & $\rho$ & 0.600 & 0.602 & 0.341 & 0.346 & 0.606 & 0.212 & 0.180 & 0.601 & 0.129 & 0.116 \\ 
		&  & $\mathrm{AUSC}$ & 0.875 & 0.868 & 0.015 & 0.014 & 0.869 & 0.010 & 0.010 & 0.869 & 0.007 & 0.007 \\ 
		\midrule
		\multirow{ 16 }{*}{ Missing thresholds } & \multirow{ 8 }{*}{ Pseudo-REML } & $\alpha_1$ & 2.000 & 1.984 & 0.121 & 0.111 & 1.986 & 0.083 & 0.081 & 1.984 & 0.053 & 0.052 \\ 
		&  & $\alpha_0$ & 1.000 & 0.994 & 0.113 & 0.106 & 0.993 & 0.078 & 0.077 & 0.992 & 0.050 & 0.050 \\ 
		&  & $\gamma_1$ & $-$2.000 & $-$1.988 & 0.065 & 0.054 & $-$1.987 & 0.044 & 0.040 & $-$1.987 & 0.027 & 0.026 \\ 
		&  & $\gamma_0$ & 1.500 & 1.494 & 0.058 & 0.047 & 1.493 & 0.039 & 0.035 & 1.492 & 0.024 & 0.023 \\ 
		&  & $\tau^2_{1}$ & 0.100 & 0.107 & 0.061 & 0.049 & 0.107 & 0.041 & 0.036 & 0.106 & 0.025 & 0.024 \\ 
		&  & $\tau^2_{0}$ & 0.100 & 0.108 & 0.060 & 0.049 & 0.109 & 0.042 & 0.037 & 0.108 & 0.026 & 0.024 \\ 
		&  & $\rho$ & 0.600 & 0.536 & 0.350 & 0.270 & 0.541 & 0.216 & 0.181 & 0.539 & 0.132 & 0.120 \\ 
		&  & $\mathrm{AUSC}$ & 0.875 & 0.872 & 0.015 & 0.014 & 0.873 & 0.011 & 0.010 & 0.873 & 0.007 & 0.007 \\ 
		\cmidrule(l){2-13}
		& \multirow{ 8 }{*}{ Riley-REML } & $\alpha_1$ & 2.000 & 1.963 & 0.120 & 0.125 & 1.965 & 0.082 & 0.080 & 1.963 & 0.052 & 0.051 \\ 
		&  & $\alpha_0$ & 1.000 & 0.988 & 0.112 & 0.112 & 0.987 & 0.076 & 0.076 & 0.986 & 0.049 & 0.049 \\ 
		&  & $\gamma_1$ & $-$2.000 & $-$1.958 & 0.067 & 0.060 & $-$1.958 & 0.045 & 0.042 & $-$1.958 & 0.028 & 0.027 \\ 
		&  & $\gamma_0$ & 1.500 & 1.459 & 0.060 & 0.053 & 1.458 & 0.041 & 0.037 & 1.458 & 0.025 & 0.024 \\ 
		&  & $\tau^2_{1}$ & 0.100 & 0.097 & 0.058 & 0.067 & 0.096 & 0.039 & 0.035 & 0.096 & 0.024 & 0.023 \\ 
		&  & $\tau^2_{0}$ & 0.100 & 0.096 & 0.057 & 0.059 & 0.096 & 0.040 & 0.035 & 0.096 & 0.025 & 0.023 \\ 
		&  & $\rho$ & 0.600 & 0.600 & 0.383 & 0.619 & 0.609 & 0.234 & 0.199 & 0.602 & 0.140 & 0.125 \\ 
		&  & $\mathrm{AUSC}$ & 0.875 & 0.870 & 0.015 & 0.016 & 0.870 & 0.011 & 0.010 & 0.870 & 0.007 & 0.007 \\ 
		\bottomrule
	\end{tabular}
	\endgroup
\end{sidewaystable}

Pseudo-REML provides almost unbiased estimators of the parameters of interest for all the examined scenarios, under a full thresholds framework, see Table \ref{tab:1}. A slightly larger standard deviation of the estimators with respect to the average standard error is not surprising as the pseudo-likelihood approach is based on a working simplified model, but such a difference reduces as the sample size $K$ or the number of thresholds increases (Table \ref{tab:2}). In presence of missing thresholds, the bias of the estimators is not affected, while standard errors slightly increase with respect to the full thresholds case. Using Riley-REML approach shows a less satisfactory behaviour. Estimators of the parameters are more biased than in the Pseudo-REML case, especially under a missing thresholds framework and when the number of thresholds increases, see, for example, the downward bias of the estimators of $\gamma_0$ and $\gamma_1$ reported in Table \ref{tab:2}. Increasing the sample size $K$ does not help in improving on the results. Similar conclusions in terms of bias and increasing standard errors are obtained when $\tau^2_{1} = \tau^2_{0} = 1$, see the Supplementary Material. 
\begin{sidewaystable}
	\centering
	\caption{Monte Carlo means (MCM), Monte Carlo standard deviations (MCSD), average of standard errors (ASE) for the parameters $\left(\alpha_0, \alpha_1, \gamma_0, \gamma_1, \tau^2_0, \tau^2_1, \rho\right)^\top$ and for AUSC, on the basis of 5,000 replicates of the simulation experiment with $\tau^2_{1} = \tau^2_{0} = 0.1, \rho = 0.6$ and increasing $K$. Maximum number of thresholds equal to 15.} 
	\label{tab:2}
	\begingroup\scriptsize
	\begin{tabular}{lllrrrrrrrrrr}
		\toprule
		& & & & \multicolumn{3}{c}{$K = 10$} & \multicolumn{3}{c}{$K = 20$} & \multicolumn{3}{c}{$K = 50$}  \\
		\cmidrule(l){5-7} \cmidrule(l){8-10}  \cmidrule(l){11-13} 
		& & & TRUE & MCM & MCSD & ASE & MCM & MCSD & ASE & MCM & MCSD & ASE \\
		\midrule
		\multirow{ 16 }{*}{ Full thresholds } & \multirow{ 8 }{*}{ Pseudo-REML } & $\alpha_1$ & 2.000 & 1.989 & 0.118 & 0.108 & 1.991 & 0.083 & 0.079 & 1.989 & 0.052 & 0.051 \\ 
		&  & $\alpha_0$ & 1.000 & 0.998 & 0.113 & 0.105 & 0.997 & 0.078 & 0.077 & 0.998 & 0.050 & 0.050 \\ 
		&  & $\gamma_1$ & $-$2.000 & $-$1.993 & 0.049 & 0.044 & $-$1.991 & 0.034 & 0.033 & $-$1.991 & 0.022 & 0.021 \\ 
		&  & $\gamma_0$ & 1.500 & 1.497 & 0.044 & 0.040 & 1.496 & 0.031 & 0.029 & 1.496 & 0.019 & 0.019 \\ 
		&  & $\tau^2_{1}$ & 0.100 & 0.121 & 0.060 & 0.047 & 0.119 & 0.041 & 0.037 & 0.120 & 0.026 & 0.025 \\ 
		&  & $\tau^2_{0}$ & 0.100 & 0.125 & 0.065 & 0.049 & 0.124 & 0.044 & 0.038 & 0.124 & 0.027 & 0.025 \\ 
		&  & $\rho$ & 0.600 & 0.463 & 0.287 & 0.208 & 0.472 & 0.193 & 0.163 & 0.481 & 0.117 & 0.109 \\ 
		&  & $\mathrm{AUSC}$ & 0.875 & 0.873 & 0.015 & 0.014 & 0.873 & 0.011 & 0.010 & 0.874 & 0.007 & 0.007 \\ 
		\cmidrule(l){2-13}
		& \multirow{ 8 }{*}{ Riley-REML } & $\alpha_1$ & 2.000 & 1.834 & 0.112 & 0.104 & 1.837 & 0.079 & 0.075 & 1.838 & 0.049 & 0.049 \\ 
		&  & $\alpha_0$ & 1.000 & 0.938 & 0.103 & 0.096 & 0.938 & 0.071 & 0.069 & 0.939 & 0.046 & 0.045 \\ 
		&  & $\gamma_1$ & $-$2.000 & $-$1.807 & 0.071 & 0.064 & $-$1.809 & 0.049 & 0.047 & $-$1.810 & 0.031 & 0.030 \\ 
		&  & $\gamma_0$ & 1.500 & 1.302 & 0.069 & 0.060 & 1.305 & 0.048 & 0.044 & 1.306 & 0.031 & 0.029 \\ 
		&  & $\tau^2_{1}$ & 0.100 & 0.086 & 0.050 & 0.042 & 0.086 & 0.034 & 0.030 & 0.086 & 0.021 & 0.020 \\ 
		&  & $\tau^2_{0}$ & 0.100 & 0.087 & 0.051 & 0.043 & 0.086 & 0.034 & 0.030 & 0.086 & 0.021 & 0.020 \\ 
		&  & $\rho$ & 0.600 & 0.601 & 0.355 & 0.335 & 0.607 & 0.222 & 0.182 & 0.610 & 0.126 & 0.117 \\ 
		&  & $\mathrm{AUSC}$ & 0.875 & 0.854 & 0.016 & 0.015 & 0.855 & 0.011 & 0.010 & 0.855 & 0.007 & 0.007 \\ 
		\midrule
		\multirow{ 16 }{*}{ Missing thresholds } & \multirow{ 8 }{*}{ Pseudo-REML } & $\alpha_1$ & 2.000 & 1.989 & 0.120 & 0.109 & 1.989 & 0.084 & 0.080 & 1.988 & 0.053 & 0.052 \\ 
		&  & $\alpha_0$ & 1.000 & 0.997 & 0.113 & 0.105 & 0.996 & 0.078 & 0.077 & 0.997 & 0.050 & 0.050 \\ 
		&  & $\gamma_1$ & $-$2.000 & $-$1.993 & 0.059 & 0.050 & $-$1.991 & 0.042 & 0.038 & $-$1.991 & 0.026 & 0.025 \\ 
		&  & $\gamma_0$ & 1.500 & 1.497 & 0.053 & 0.044 & 1.495 & 0.037 & 0.034 & 1.496 & 0.023 & 0.023 \\ 
		&  & $\tau^2_{1}$ & 0.100 & 0.116 & 0.060 & 0.048 & 0.115 & 0.042 & 0.037 & 0.116 & 0.026 & 0.025 \\ 
		&  & $\tau^2_{0}$ & 0.100 & 0.120 & 0.065 & 0.050 & 0.119 & 0.043 & 0.038 & 0.119 & 0.027 & 0.025 \\ 
		&  & $\rho$ & 0.600 & 0.482 & 0.306 & 0.220 & 0.489 & 0.204 & 0.170 & 0.498 & 0.121 & 0.113 \\ 
		&  & $\mathrm{AUSC}$ & 0.875 & 0.873 & 0.015 & 0.014 & 0.873 & 0.011 & 0.010 & 0.873 & 0.007 & 0.007 \\ 
		\cmidrule(l){2-13}
		& \multirow{ 8 }{*}{ Riley-REML } & $\alpha_1$ & 2.000 & 1.901 & 0.116 & 0.110 & 1.902 & 0.081 & 0.077 & 1.902 & 0.051 & 0.050 \\ 
		&  & $\alpha_0$ & 1.000 & 0.971 & 0.108 & 0.104 & 0.970 & 0.074 & 0.073 & 0.970 & 0.048 & 0.047 \\ 
		&  & $\gamma_1$ & $-$2.000 & $-$1.876 & 0.072 & 0.063 & $-$1.876 & 0.049 & 0.046 & $-$1.877 & 0.031 & 0.030 \\ 
		&  & $\gamma_0$ & 1.500 & 1.373 & 0.066 & 0.059 & 1.374 & 0.047 & 0.043 & 1.375 & 0.029 & 0.028 \\ 
		&  & $\tau^2_{1}$ & 0.100 & 0.092 & 0.054 & 0.047 & 0.092 & 0.037 & 0.032 & 0.092 & 0.023 & 0.022 \\ 
		&  & $\tau^2_{0}$ & 0.100 & 0.093 & 0.056 & 0.050 & 0.092 & 0.037 & 0.033 & 0.092 & 0.023 & 0.022 \\ 
		&  & $\rho$ & 0.600 & 0.595 & 0.369 & 0.439 & 0.600 & 0.231 & 0.189 & 0.603 & 0.130 & 0.122 \\ 
		&  & $\mathrm{AUSC}$ & 0.875 & 0.863 & 0.015 & 0.015 & 0.863 & 0.011 & 0.010 & 0.864 & 0.007 & 0.007 \\ 
		\bottomrule
	\end{tabular}
	\endgroup
\end{sidewaystable}

When evaluating the performance of the methods in terms of empirical coverage for confidence intervals for the regression coefficients and the AUSC at 95\% nominal level, the pseudo-likelihood solution provides satisfactory results, either when missing thresholds are present or not, see Figures \ref{fig:1} and \ref{fig:2}. As expected from a theoretical point of view, results improve as the sample size increases, while they are not affected by the number of thresholds. The use of REML or the ML estimation methods has no relevant effects on the results. Inference based on Riley et al. \cite{riley2014meta} approach, instead, provides unreliable conclusions. Empirical coverages of confidence intervals tend to be dramatically lower from the target 95\% level when the number of thresholds increases, especially in case of full thresholds per study, see Figure \ref{fig:1}. Increasing the sample size does not improve on the conclusion. The worst effects are experienced when the interest is on the regression coefficients $\gamma_0$ and $\gamma_1$, with empirical coverages approaching zero. As for pseudo-likelihood approach, REML and ML estimation techniques provide similar conclusions. Unsatisfactory results with increasing number of thresholds are not surprising: the largest the number of thresholds used to subdivide the observations, the smaller the amount of information available for inference between consecutive thresholds. Less information useful to properly estimate the model components leads to large bias and large variability, with consequent poor empirical coverage probabilities. A similar behaviour is obtained for scenarios with different values of between-study variances, see the Supplementary Material. 

Under different values of $\rho$, there is no substantial variation in terms of bias and standard error of the estimators from either the pseudo-likelihood approach and the approach in Riley et al. \cite{riley2014meta}, see Tables in the Supplementary Material. Similarly, the relative performance of the methods in terms of empirical coverage for Wald-type confidence intervals is maintained, see Figures in the Supplementary Material. 

Substantial differences between the competing approaches occur in terms of failure rate of the estimation process. Convergence problems affect the estimation of the model in Riley et al. \cite{riley2014meta}, typically with variance/covariance parameters on the boundary of the parameter space. Convergence issues are particularly relevant for large number of thresholds per study, large heterogeneity and no missing data. Results indicate a failure rate exceeding 20\% under small heterogeneity and exceeding 26\% for 50 studies and 20 thresholds, under large heterogeneity, for $\rho=0.6$. Increasing $\rho$ leads to larger failure rate, up to 50\%, when the between-study heterogeneity $\tau^2_0, \tau^2_1$ and the sample size $K$ are small. Conversely, convergence problems have been very rarely encountered when applying the pseudo-likelihood approach, with a maximum failure rate around 3.5\% in case of small number of thresholds ($m=3$) and small sample size ($K=10$). The failure rate becomes zero when increasing the number of studies and the number of thresholds per study.

\section{The spot protein to creatinine ratio data}
\begin{figure}[h]
	\begin{center}
		\includegraphics[width=3.5in]{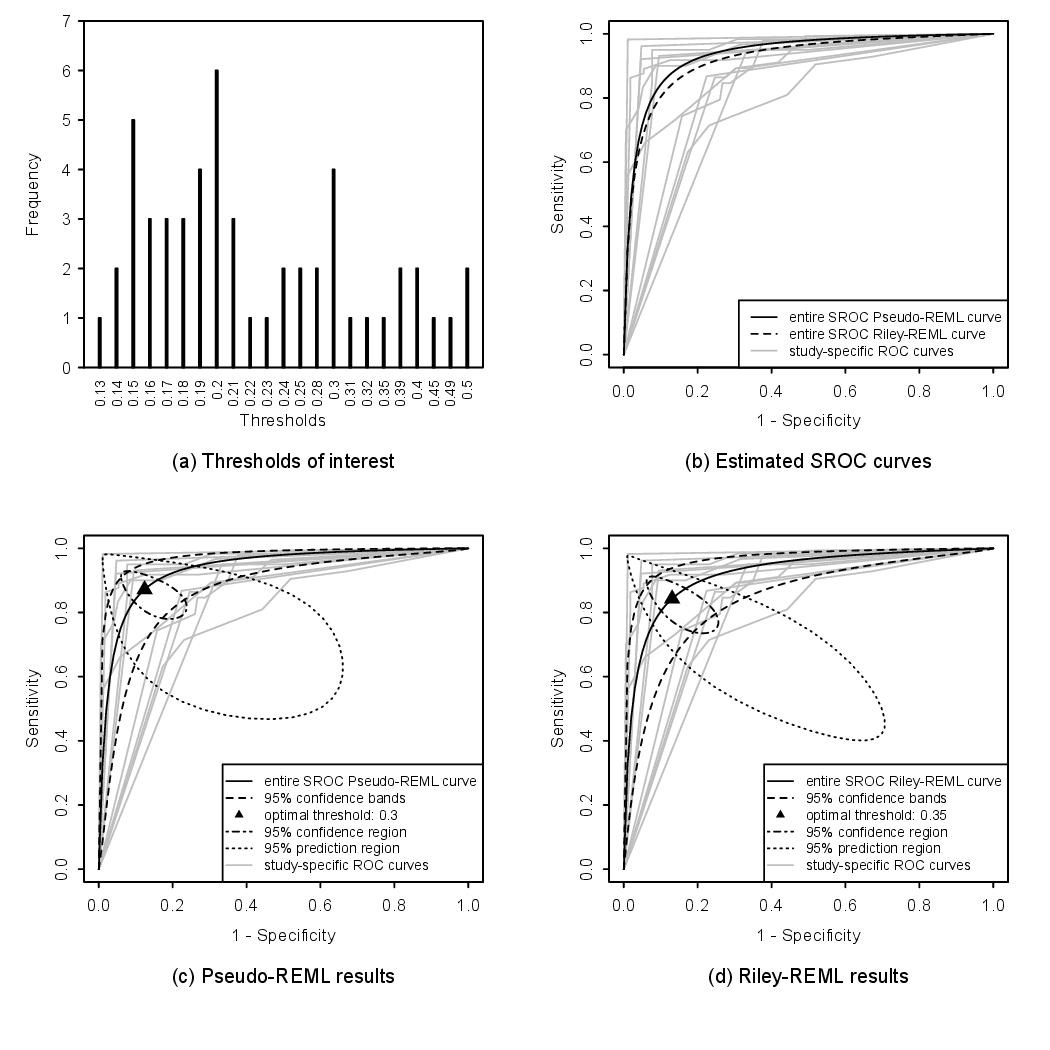}
	\end{center}
	\caption{Spot PCR data. Panel (a): frequency of the thresholds. Panel (b): study-specific and estimated SROC curves. Panels (c)--(d): SROC curve with 95\% confidence bands; optimal sensitivity and 1-specificity with 95\% confidence region; prediction region. Methods: Riley-REML and Pseudo-REML.}
	\label{fig:PCR}
\end{figure}

Significant proteinuria (equal or larger than 0.3 g per 24 hour) after the 20-th week of gestation is relevant for the diagnosis of pre-eclampsia, a progressive disorder characterized by high blood pressure. The spot protein to creatinine ratio (PCR) is frequently used as a diagnostic test for detecting pathology and it represents an alternative to the time consuming and discomforting reference standard given by the 24 hour urine collection. Within this framework, Morris et al. \cite{morris2012diagnostic} perform a meta-analysis based on 13 studies with 23 different thresholds ranging from 0.13 to 0.5, in ratio. The data are available in the Supplementary Material. Five studies consider only one threshold, while the remaining studies report a number of thresholds from 2 to 9, see Figure \ref{fig:PCR}(a) for details. Characteristics of the studies included in the meta-analysis are available in Riley et al. \cite{riley2014meta}, where the results from the multivariate normal model described in Section \ref{sec:bgs:2} are reported. The missing threshold problem is faced under the missing completely at random assumption. 
\begin{table}[htbp]
	\begin{center}
		\caption{Spot PCR data. Results for the parameters in model (\ref{md1}) and for AUSC from using Pseudo-REML and Riley-REML.}
		\label{tab:pcr}
		\begin{tabular}{l r r r | r r r}
			\toprule
			& \multicolumn{3}{c|}{Pseudo-REML} & \multicolumn{3}{c}{Riley-REML} \\
			\midrule
			& Estimate & Std. error & $p$-value & Estimate & Std. error & $p$-value \\ 
			\midrule
			$\alpha_1$ & 3.555 & 0.334 & $<$ 0.001 & 2.884 & 0.407 & $<$ 0.001 \\
			$\alpha_0$ & 0.327 & 0.531 & 0.538 & 0.557 & 0.410 & 0.175 \\
			$\gamma_1$ & $-$5.456 & 0.814 & $<$ 0.001 & $-$3.437 & 1.123 & 0.002 \\
			$\gamma_0$ & 5.427 & 1.779 & 0.002 & 3.819 & 0.976 & $<$ 0.001  \\
			$\tau_1^2$ & 0.629 & 0.399 & -- & 0.652 & 0.414 & -- \\
			$\tau_0^2$ & 1.045 & 0.558 & -- & 1.171 & 0.670 & -- \\
			$\rho$ & 0.682 & 0.247 & -- & 0.915 & 0.094 & -- \\
			\midrule[1.5pt]
			& Estimate & Std. error & 95\% interval& Estimate & Std. error & 95\% interval \\
			\midrule
			AUSC & 0.937 & 0.033 & (0.831, 0.978) & 0.922 & 0.035 & (0.821, 0.968) \\
			\bottomrule
		\end{tabular}
	\end{center}
\end{table}

Table \ref{tab:pcr} reports estimates, sandwich estimates of standard errors and p-values for the test of significance of the parameters of interest using Riley et al. \cite{riley2014meta} approach and the pseudo-likelihood approach, both under REML estimation. Regression coefficients apart from $\alpha_0$ are significantly different from zero according to both the approaches. The large estimated between-study variances suggest the presence of substantial heterogeneity between the studies.

The SROC curves from Pseudo-REML and Riley-REML are reported in Figure \ref{fig:PCR}(b). The resulting values of $\gamma_0$ and $\gamma_1$ allow to conclude for the concavity of the SROC curve, see the discussion in Section~\ref{sec:proposal_pseudo}. As the SROC curve from Pseudo-REML lies above the Riley-REML alternative, the estimated AUC is larger, namely, it is equal to 0.937 against 0.922 from Riley-REML, see Table \ref{tab:pcr}. Both the values are considerably greater than the AUSC equal to 0.69 provided in Morris et al. \cite{morris2012diagnostic}, although it is worth mentioning that the area underneath the SROC curve in Morris et al. \cite{morris2012diagnostic} seems to be higher than reported. Basing on the Youden index \cite{youden1950index}, the optimal threshold obtained from Pseudo-REML is 0.3, while the one obtained from Riley-REML is 0.35. The corresponding summary sensitivity and summary specificity are 87.2\% and 87.6\%, respectively, for Pseudo-REML and 84.3\% and 86.9\%, respectively, for Riley-REML. The values are reported into the ROC space in Figure \ref{fig:PCR}(c)-(d) for Pseudo-REML and Riley-REML, respectively, together with the associated 95\% confidence regions based on the normal approximation. The 95\% prediction regions associated to predicted sensitivity and predicted specificity at a given threshold are drawn as well. See Higgins et al. \cite{higgins2009re}, Riley et al. \cite{riley2011interpretation}, Gasparrini et al. \cite{gasparrini2012} for details. In the same graph, the 95\% confidence bands of the SROC curves are superimposed.

\section{Discussion} 
This paper proposed a pseudo-likelihood strategy for inference in meta-analysis of test accuracy studies when test performance is evaluated at multiple thresholds. Inspired by the idea of composite likelihood, the method allows a simplified specification of the within-study model with respect to the proposal in Riley et al. \cite{riley2014meta}, as it is developed under a working independence assumption between sensitivities and between specificities at different thresholds in the same study. Accordingly, the information about within-study covariances, which is typically of difficult specification or estimation, is not required. The lack of substantial convergence issues and the straightforward implementation of the approach using standard software are further appealing features of the method. In the meanwhile, the pseudo-likelihood method maintains advantages of the approach in Riley et al. \cite{riley2014meta}, such as the possibility to account for studies with one single threshold as well as studies with missing thresholds and the monotonicity of sensitivity and specificity with increasing thresholds. 

Simulation studies under different scenarios show a satisfactory behaviour in terms of bias, standard error and empirical coverage probabilities for the estimators of the parameter of interest, in this way largely improving on the results from the methodology in Riley et al. \cite{riley2014meta}. Such a performance is maintained under varying sample size, number of thresholds, between-study heterogeneity or correlation as well as the presence of missing thresholds. When the number of thresholds increases, the reduced information available between consecutive thresholds deeply affects the estimates of the within-study covariances as developed in Riley et al. \cite{riley2014meta}, with consequent poor results. Conclusions from the pseudo-likelihood approach, instead, are not affected.

The pseudo-likelihood approach has been implemented in the \texttt R \cite{r_software} programming language. The code is available upon request from the authors. The application of the method is illustrated on an additional dataset about a test for identification of type 2 diabetes mellitus. The results are reported in the Supplementary Material.

Although we did not account for the presence of additional study-specific covariates in the model, such an extension is straightforward, in the simplest way by modifying the specification of the mean in (\ref{md1}) to include covariates information.
Further extensions, also suggested in Riley et al. \cite{riley2014meta}, are expected to provide no substantial obstacles to the application of the pseudo-likelihood approach. We refer to different specifications of the model, including the presence of random slopes together with random intercepts or more complex mean specification as functions of the thresholds, e.g. polynomial terms. Deeper investigations in this direction represent an interesting line of future research. 

The presence of missing information about test accuracy at different thresholds is a relevant and common issue in medical investigations. The method in Riley et al. \cite{riley2014meta} and the pseudo-likelihood proposal allow to account for missing completely at random thresholds. Such an assumption can be violated, for example, in case of selective reporting of thresholds, where thresholds are more likely to be reported when they are associated with large values of sensitivity and specificity (\cite{korevaar2014, riley2015meta, ensor2017meta}). Further work is needed on the development of the pseudo-likelihood approach in order to address the different features of the missing thresholds problem.

\section*{Acknowledgments}
This work was supported by a grant from the University of Padova (Progetti di Ricerca di Ateneo 2015, CPDA153257).
The authors are grateful to Ruggero Bellio for discussion and comments on a preprint version of the manuscript.

\section*{Supporting information}
The Supplementary Material provides $i)$ details about the computation of the SROC curve and the area underneath, $ii)$ the results of additional simulation studies performed within the scenarios described in Section 4, $iii)$ the spot protein to creatinine ratio data and $iv)$ the results of the application of the pseudo-likelihood approach to an additional real dataset. At the time of writing, the \texttt R code to apply the pseudo-likelihood approach to the datasets is available upon request from the authors.

\newpage
\begin{center}\LARGE{
Supplementary Material for \\
"A pseudo-likelihood approach for multivariate meta-analysis of test accuracy studies with multiple thresholds"
\\
by \\
Annamaria Guolo and Duc Khanh To}
\end{center}

\vspace{2cm}

\appendix
\setcounter{figure}{0} \renewcommand{\thefigure}{S.\arabic{figure}}
\setcounter{table}{0} \renewcommand{\thetable}{S.\arabic{table}}

\section{The SROC curve and the area underneath} 
According to model specification in Section 2 of the main paper, the summary sensitivity and specificity can be obtained as
\begin{eqnarray}
SSe_j &=& \logit^{-1}\left(\alpha_1 + \gamma_1 x_j\right) = \Pro\left(T \ge x_j | D = 1\right) \nonumber, \\
SSp_j &=& \logit^{-1}\left(\alpha_0 + \gamma_0 x_j\right) = \Pro\left(T < x_j | D = 0\right) \nonumber,
\end{eqnarray}
where $D$ is the true disease status of a patient verified by the reference standard, with $D = 1$ for a diseased patient and $D = 0$ for a nondiseased patient. Let $S_{D, \alpha_1, \gamma_1}(x) = \Pro\left(T \ge x | D = 1\right)$ and $S_{\bar{D}, \alpha_0, \gamma_0}(x) = \Pro\left(T \ge x | D = 0\right)$. Then, the summary ROC curve can be represented as
\[
\mathrm{SROC}\left(t;\alpha_1, \alpha_0, \gamma_1, \gamma_0\right) = S_{D, \alpha_1, \gamma_1}\left(S^{-1}_{\bar{D}, \alpha_0, \gamma_0}(t)\right), \qquad t \in (0,1),
\]
see \cite{pep2}. For given $t \in (0,1)$, 
$$
S^{-1}_{\bar{D}, \alpha_0, \gamma_0}(t) = \frac{\logit(1 - t) - \alpha_0}{\gamma_0} = x.
$$
The associated variance is obtained using the delta method,
\begin{eqnarray}
\lefteqn{\V\left\{\mathrm{SROC}\left(t;\alpha_1, \alpha_0, \gamma_1, \gamma_0\right)\right\}} \nonumber \\
&=& \frac{\partial}{\partial \boldsymbol\beta} \mathrm{SROC}\left(t;\alpha_1, \alpha_0, \gamma_1, \gamma_0\right) \V\left(\boldsymbol\beta\right)\frac{\partial^\top}{\partial \boldsymbol\beta^\top} \mathrm{SROC}\left(t;\alpha_1, \alpha_0, \gamma_1, \gamma_0\right), \nonumber
\end{eqnarray}
where the first derivative of SROC with respect to the components of $\boldsymbol \beta$ is
\begin{eqnarray}
\lefteqn{\frac{\partial}{\partial \alpha_1} \mathrm{SROC}\left(t;\alpha_1, \alpha_0, \gamma_1, \gamma_0\right)}  \nonumber \\
&=& \mathrm{SROC}\left(t;\alpha_1, \alpha_0, \gamma_1, \gamma_0\right) \left\{1 - \mathrm{SROC}\left(t;\alpha_1, \alpha_0, \gamma_1, \gamma_0\right)\right\}, \nonumber \\
\lefteqn{\frac{\partial}{\partial \alpha_0} \mathrm{SROC}\left(t;\alpha_1, \alpha_0, \gamma_1, \gamma_0\right)} \nonumber \\
&=& - \frac{\gamma_1}{\gamma_0} \mathrm{SROC}\left(t;\alpha_1, \alpha_0, \gamma_1, \gamma_0\right) \left\{1 - \mathrm{SROC}\left(t;\alpha_1, \alpha_0, \gamma_1, \gamma_0\right)\right\}, \nonumber \\
\lefteqn{\frac{\partial}{\partial \gamma_1} \mathrm{SROC}\left(t;\alpha_1, \alpha_0, \gamma_1, \gamma_0\right)} \nonumber \\
&=& \frac{\logit(1 - t) - \alpha_0}{\gamma_0} \mathrm{SROC}\left(t;\alpha_1, \alpha_0, \gamma_1, \gamma_0\right) \left\{1 - \mathrm{SROC}\left(t;\alpha_1, \alpha_0, \gamma_1, \gamma_0\right)\right\}, \nonumber \\
\lefteqn{\frac{\partial}{\partial \gamma_0} \mathrm{SROC}\left(t;\alpha_1, \alpha_0, \gamma_1, \gamma_0\right)} \nonumber \\
&=& -\frac{\gamma_1}{\gamma^2_0} \left\{\logit(1 - t) - \alpha_0\right\} \mathrm{SROC}(t;\alpha_1, \alpha_0, \gamma_1, \gamma_0) \left\{1 - \mathrm{SROC}\left(t;\alpha_1, \alpha_0, \gamma_1, \gamma_0\right)\right\}. \nonumber
\end{eqnarray}

The estimate of AUSC and the associated variance can be obtained by exploiting the functional form of the SROC curve (\cite{pep2}) 
\begin{eqnarray}\nonumber
\mathrm{AUSC}\left(\alpha_1, \alpha_0, \gamma_1, \gamma_0\right)  &=&\int_{0}^{1} \mathrm{SROC}\left(t;\alpha_1, \alpha_0, \gamma_1, \gamma_0\right) \ud t\\ \nonumber
&=& \int_{0}^{1} \logit^{-1} \left\{\alpha_1 + \gamma_1 \frac{\logit(1 - t) - \alpha_0}{\gamma_0} \right\} \ud t.
\end{eqnarray}
Similarly to the SROC curve, the variance of AUSC can be obtained by delta method
\[
\frac{\partial}{\partial \boldsymbol \beta}\mathrm{AUSC}\left(\alpha_1, \alpha_0, \gamma_1, \gamma_0\right) = \int_{0}^{1}\frac{\partial}{\partial \boldsymbol \beta} \mathrm{SROC}\left(t;\alpha_1, \alpha_0, \gamma_1, \gamma_0\right) \ud t.
\]

\section{Concavity of the SROC curve} 
Consider the SROC curve following the general formulation 
$$
f (t; a, b) = \logit^{-1} \{a + b \ \logit(t)\},  \ \ t \in (0, 1), \ a, b \in \Real .
$$ 
The concavity of $f (t; a, b)$ is guaranteed when $b=1$. For $0 < b < 1$, the function is concave at the top-right corner of the SROC, where $t$ is close to 1, while for $b > 1$ the function is concave at the bottom-left corner of the SROC, where $t$ is close to 0. 
The SROC curve proposed in Section 2 of the main paper takes the same form as the function $f (t; a, b)$ with $a = \alpha_1 - \gamma_1 \alpha_0/\gamma_0$ and $b=-\gamma_1/\gamma_0$. The concavity properties of the proposed SROC curve can be derived accordingly.

The proposed SROC curve can be re-written in terms of the SROC curve in Moses et al. \cite{moses1993combining2}, that is based on the relationship $D = a + bS$, where
$$
 D= \log{\frac{SE}{1-SE}} - \log{\frac{1-SP}{SP}}
 $$
 is equivalent to the diagnostic odds ratio and
 $$
 S=\log{\frac{SE}{1-SE}} + \log{\frac{1-SP}{SP}} 
 $$
 can be interpreted as a measure of diagnostic threshold. After estimating $a$ and $b$ through, for example, weighted least squares criterion, with weights given by the inverse of the variance of $D$, the SROC curve can be drawn as
 \begin{equation}\label{eqn:morris}
SE = \frac{
e^{\frac{a}{1-b}} \left(\frac{1-SP}{SP}\right)^{\left(\frac{1+b}{1-b}\right)}}{ 1 + e^{\frac{a}{1-b}} \left(\frac{1-SP}{SP}\right)^{\left(\frac{1+b}{1-b}\right)}
}.
\end{equation}
Expression (\ref{eqn:morris}) can be re-written as 
\begin{eqnarray}
SE &=& \frac{\exp\left(\frac{a}{1-b}\right) \exp \left(\frac{1+b}{1-b}\log\left(\frac{1-SP}{SP}\right)\right)}{1 + \exp\left(\frac{a}{1-b}\right) \exp \left(\frac{1+b}{1-b}\log\left(\frac{1-SP}{SP}\right)\right)} \nonumber\\
&=&  \logit^{-1} \left\{\frac{a}{1-b} + \frac{1+b}{1-b}\logit(t)\right\}, \nonumber
\end{eqnarray}
where $t = 1 - SP$ and $-1 < b < 1$. 
 It is easy to see that
\[
\left\{
\begin{array}{r c l}
\alpha_1 - \dfrac{\gamma_1}{\gamma_0} \alpha_0 & = &\dfrac{a}{1-b}, \\ [8pt]
- \dfrac{\gamma_1}{\gamma_0} & = &\dfrac{1+b}{1-b},
\end{array}
\right.
\]
or, equivalently,
\[
\left\{
\begin{array}{r c l}
a & = & 2 \dfrac{\alpha_1\gamma_0 - \alpha_0\gamma_1}{\gamma_0 - \gamma_1} , \\ [8pt]
b & = & \dfrac{\gamma_1 + \gamma_0}{\gamma_1 - \gamma_0}.
\end{array}
\right.
\]
As Walter \cite{walter20022} illustrates, the SROC curve is symmetric (with respect to the line $1 - t$, for $t \in (0,1)$) if $a > 0$ and $b = 0$. For given $a > 0$, the SROC curve is convex at the top-right of corner if $-1 < b < 0$ and at the bottom-left of corner if $0 < b < 1$. Analogously, the proposed SROC curve is concave if $\gamma_1 = -\gamma_0$, and convex if $-1 < \dfrac{\gamma_1 + \gamma_0}{\gamma_1 - \gamma_0}< 0$ or $0 < \dfrac{\gamma_1 + \gamma_0}{\gamma_1 - \gamma_0} < 1$ in case $\alpha_1 - \alpha_0\dfrac{\gamma_1}{\gamma_0} > 0$.

Again, the proposed SROC curve can be re-written in terms of the SROC curve in Arends et al. \cite{arends2008bivariate2}, this is defined as  can be rewritten as
\begin{equation}
SE = \logit^{-1} \left\{c + d\ \logit(t)\right\},
\end{equation}
with $t = 1 - SP$. Comparing the expression with that of the proposed SROC curve
\[
\left\{
\begin{array}{r c l}
c & = & \alpha_1 - \dfrac{\gamma_1}{\gamma_0} \alpha_0, \\ [8pt]
d & = & -\dfrac{\gamma_1}{\gamma_0}.
\end{array}
\right.
\]
Then, the SROC curve is concave when $d = 1$ and it is convex when $0 < d < 1$ or $d > 1$.

\section{Additional simulation results}
\begin{figure}[hbtp]
	\begin{center}
		\includegraphics[width=3.5in]{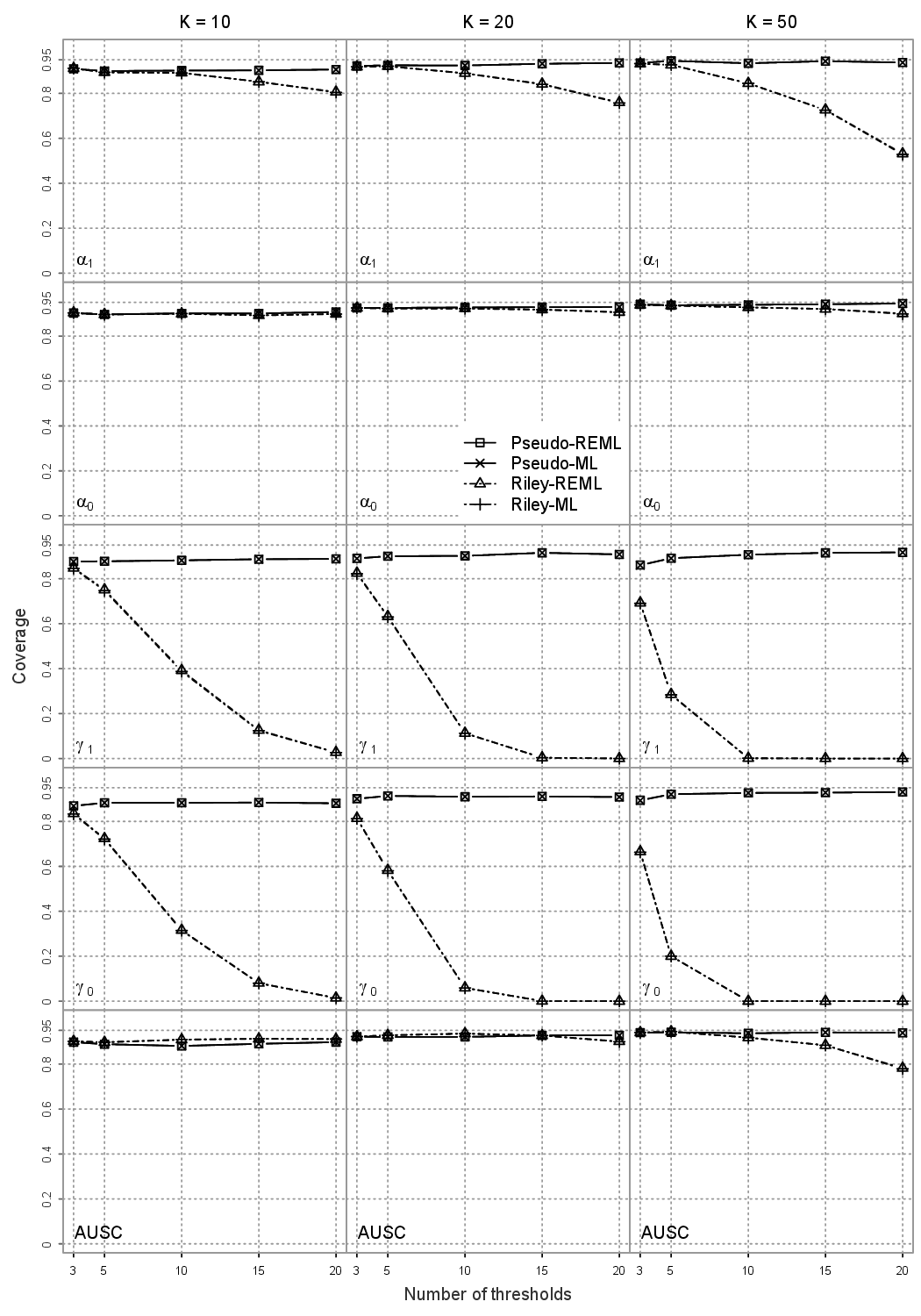}
	\end{center}
	\caption{Empirical coverages of confidence intervals at nominal level 0.95 for $\alpha_0, \alpha_1, \gamma_0, \gamma_1$ and AUSC using Pseudo-REML, Pseudo-ML, Riley-REML and Riley-ML approaches, for $\tau^2_{1} = \tau^2_{0} = 1, \rho = 0.6$ and increasing $K$. All thresholds of interest are reported in the studies.}
	\label{fig:S1}
\end{figure}

\begin{figure}[h]
	\begin{center}
		\includegraphics[width=3.5in]{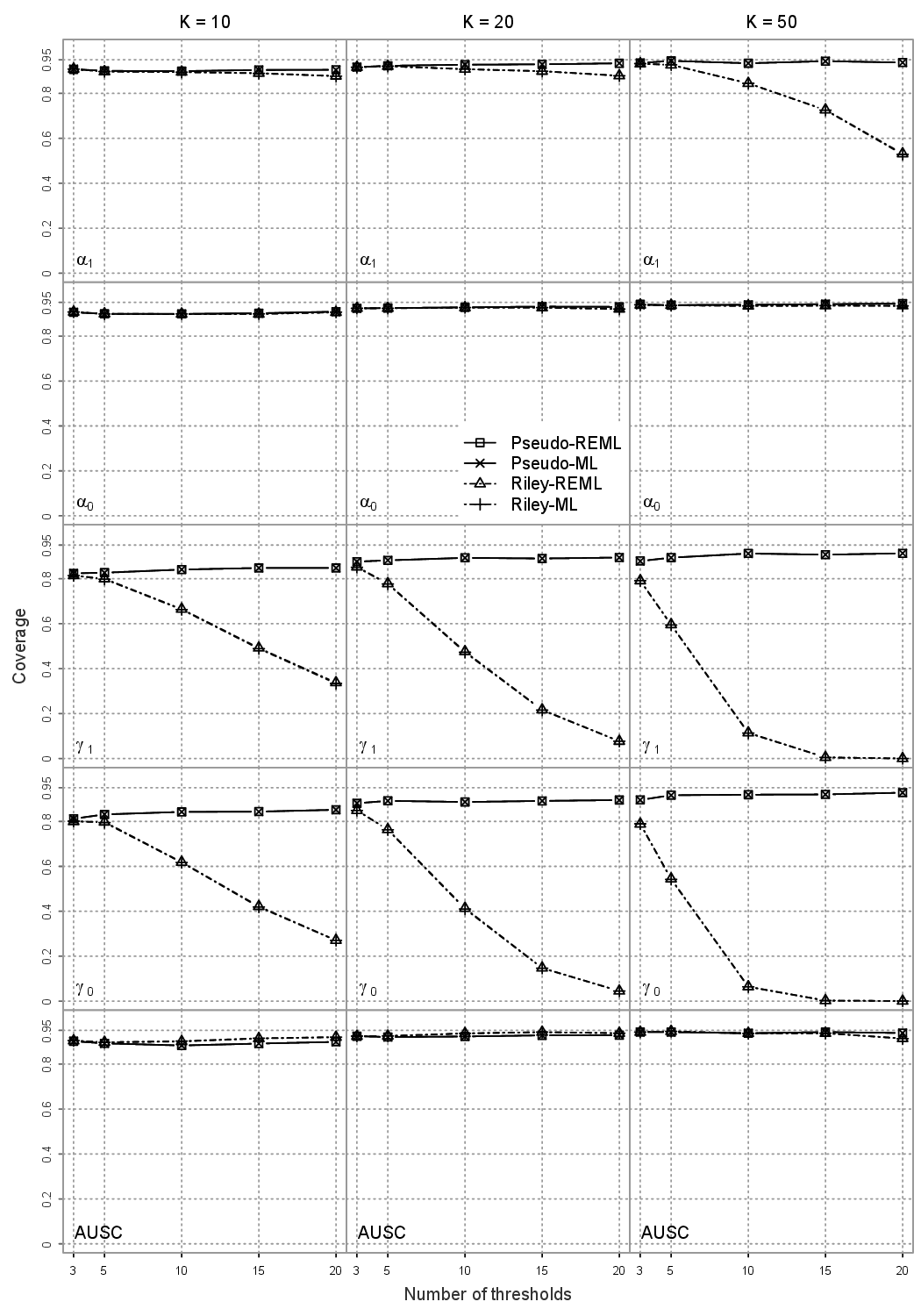}
		\caption{Empirical coverages of confidence intervals at nominal level 0.95 for $\alpha_0, \alpha_1, \gamma_0, \gamma_1$ and AUSC using Pseudo-REML, Pseudo-ML, Riley-REML and Riley-ML approaches, for $\tau^2_{1} = \tau^2_{0} = 1, \rho = 0.6$ and increasing $K$. Different thresholds of interest are reported in the studies.}
			\label{fig:S2}
	\end{center}
\end{figure}

\begin{figure}[hbtp]
	\begin{center}
		\includegraphics[width=3.5in]{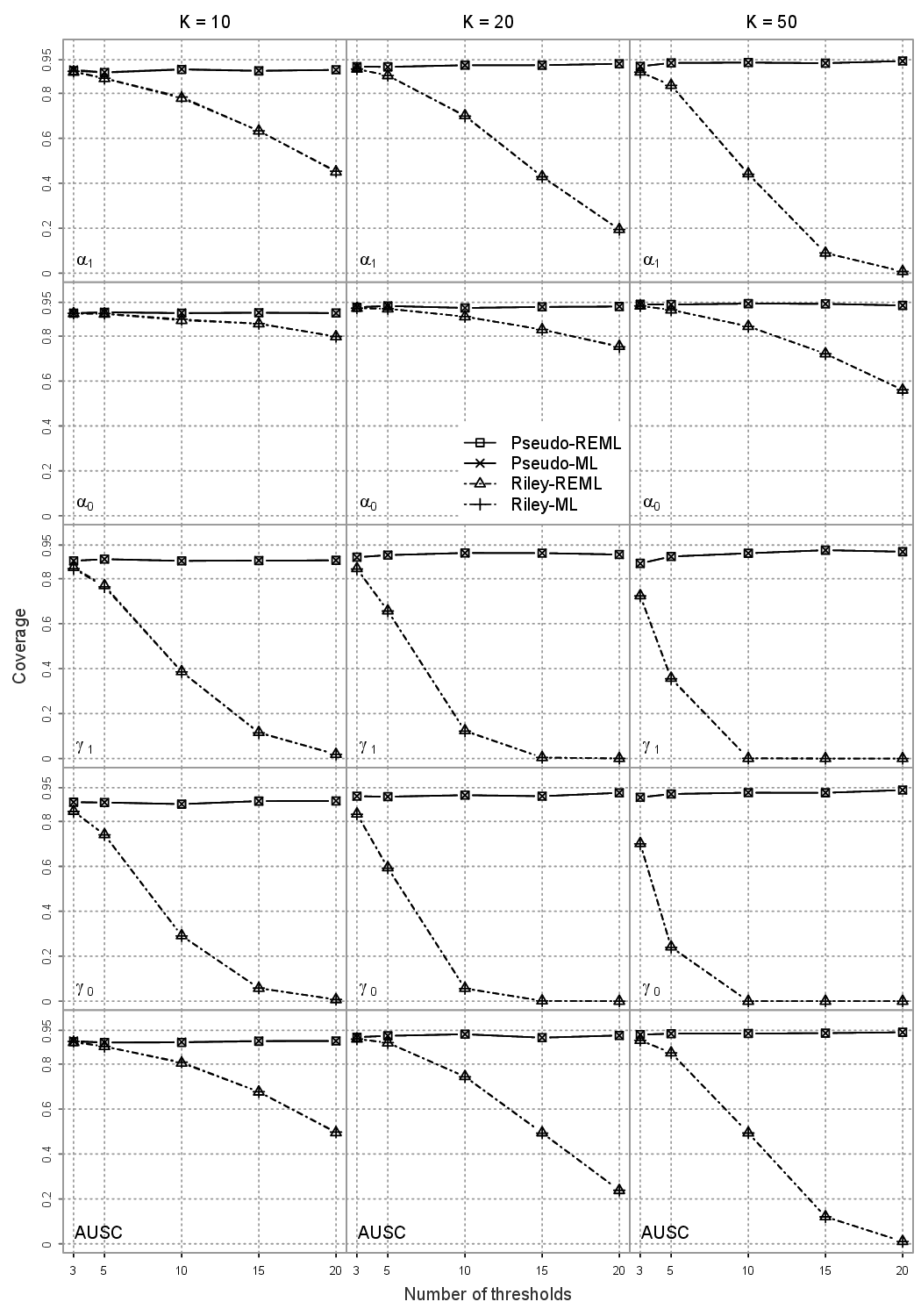}
	\end{center}
	\caption{Empirical coverages of confidence intervals at nominal level 0.95 for $\alpha_0, \alpha_1, \gamma_0, \gamma_1$ and AUSC using Pseudo-REML, Pseudo-ML, Riley-REML and Riley-ML approaches, for $\tau^2_{1} = \tau^2_{0} = 0.1, \rho = 0.3$ and increasing $K$. All thresholds of interest are reported in the studies.}
	\label{fig:S3}
\end{figure}

\begin{figure}[h]
	\begin{center}
		\includegraphics[width=3.5in]{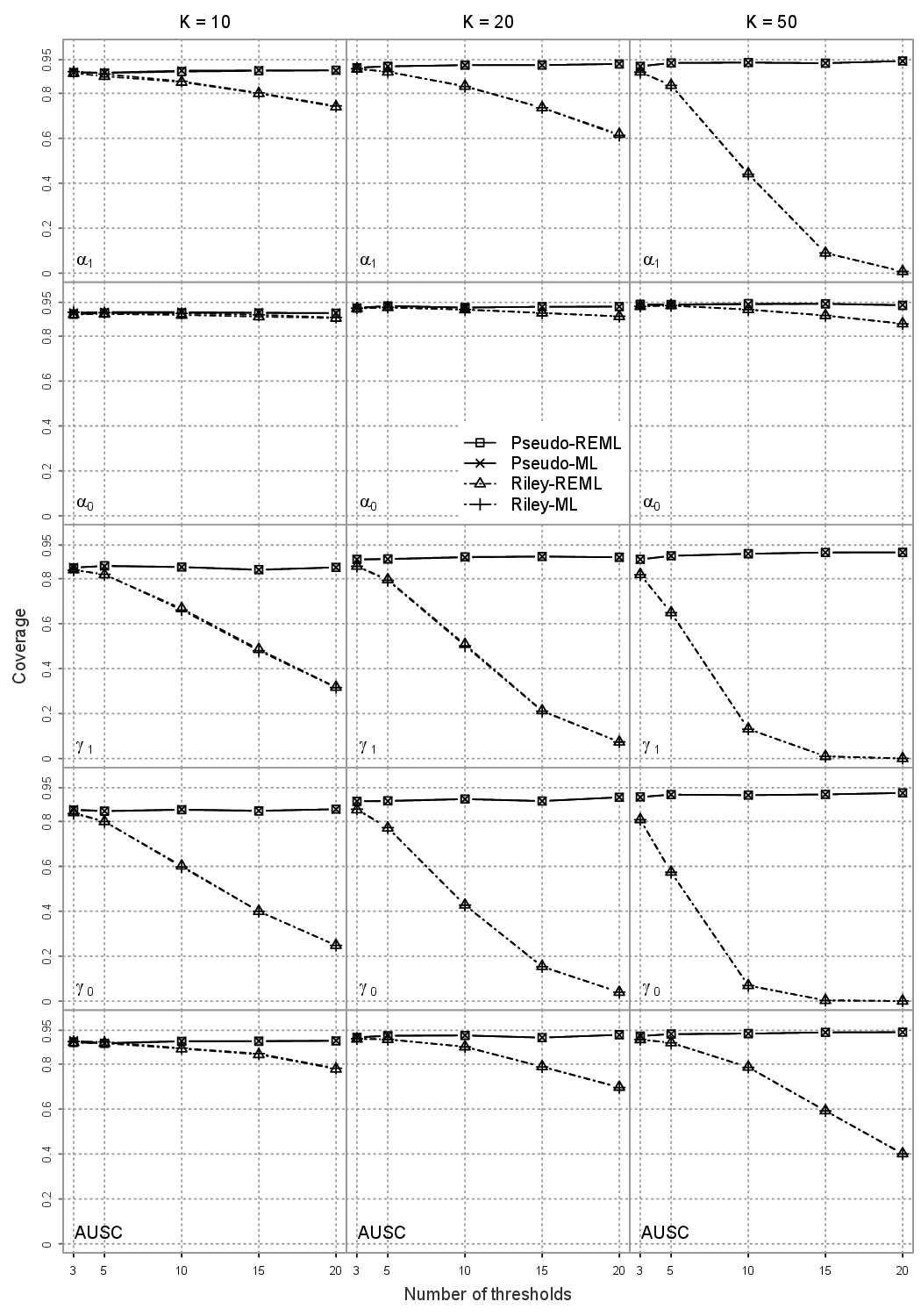}
		\caption{Empirical coverages of confidence intervals at nominal level 0.95 for $\alpha_0, \alpha_1, \gamma_0, \gamma_1$ and AUSC using Pseudo-REML, Pseudo-ML, Riley-REML and Riley-ML approaches, for $\tau^2_{1} = \tau^2_{0} = 0.1, \rho = 0.3$ and increasing $K$. Different thresholds of interest are reported in the studies.}
			\label{fig:S4}
	\end{center}
\end{figure}

\begin{figure}[h]
	\begin{center}
		\includegraphics[width=3.5in]{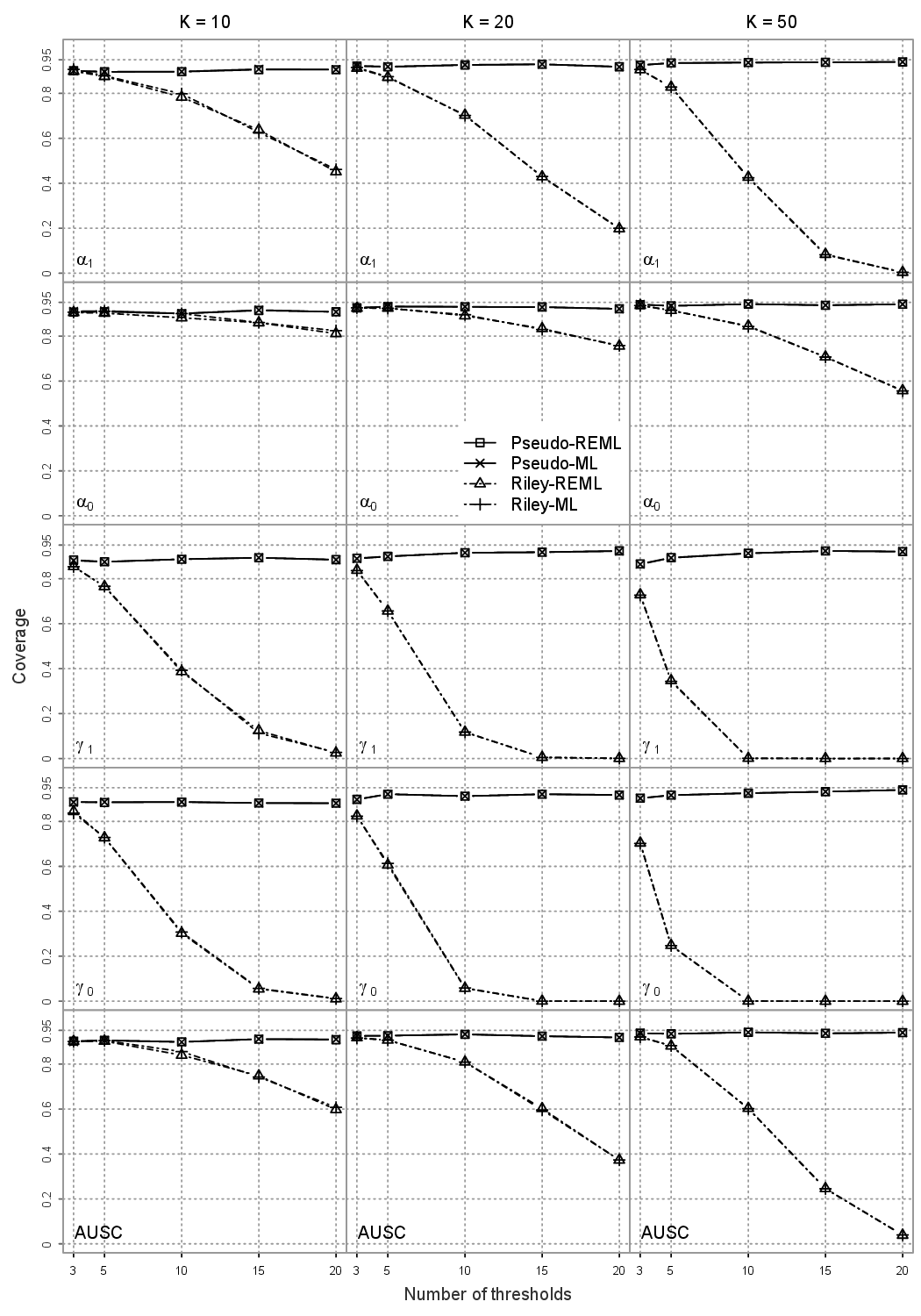}
	\end{center}
	\caption{Empirical coverages of confidence intervals at nominal level 0.95 for $\alpha_0, \alpha_1, \gamma_0, \gamma_1$ and AUSC using Pseudo-REML, Pseudo-ML, Riley-REML and Riley-ML approaches, for $\tau^2_{1} = \tau^2_{0} = 0.1, \rho = 0.9$ and increasing $K$. All thresholds of interest are reported in the studies.}
	\label{fig:S5}
\end{figure}

\begin{figure}[h]
	\begin{center}
		\includegraphics[width=3.5in]{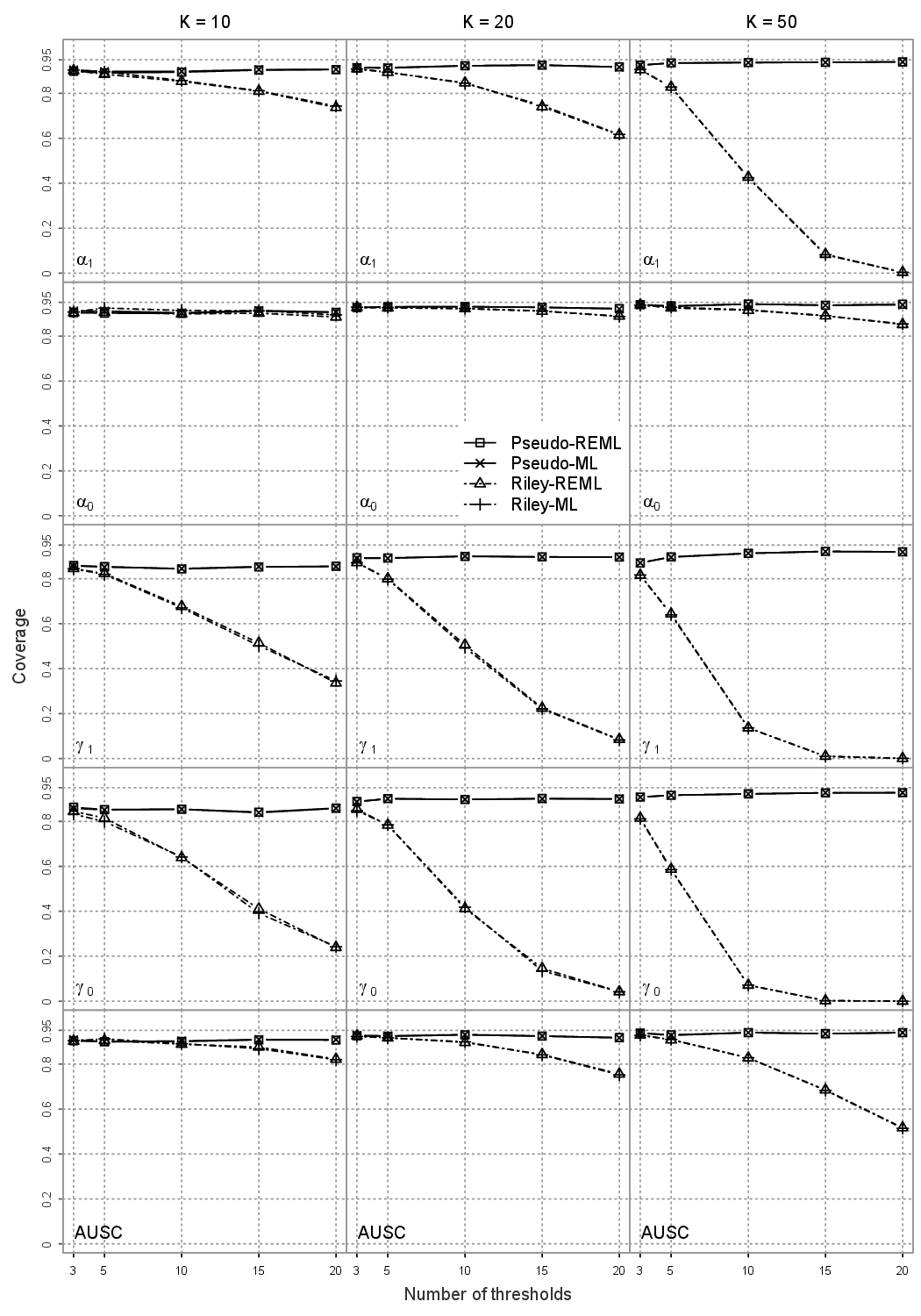}
	\end{center}
	\caption{Empirical coverages of confidence intervals at nominal level 0.95 for $\alpha_0, \alpha_1, \gamma_0, \gamma_1$ and AUSC using Pseudo-REML, Pseudo-ML, Riley-REML and Riley-ML approaches, for $\tau^2_{1} = \tau^2_{0} = 0.1, \rho = 0.9$ and increasing $K$. Different thresholds of interest are reported in the studies.}
	\label{fig:S6}
\end{figure}

This section contains the results of additional simulation studies performed within the scenarios described in Section 4 of the main paper, namely,
\begin{itemize}
\item Monte Carlo mean, Monte Carlo standard deviation, average of standard errors for parameters $\left(\alpha_0, \alpha_1, \gamma_0, \gamma_1, \tau^2_0, \tau^2_1, \rho\right)^\top$ and for AUSC using Pseudo-REML and Riley-REML when $\rho=0.6$ and $\tau^2_0=\tau^2_1=1.0$ (Tables \ref{tab:S1}-\ref{tab:S2});
\item Monte Carlo mean, Monte Carlo standard deviation, average of standard errors for parameters $\left(\alpha_0, \alpha_1, \gamma_0, \gamma_1, \tau^2_0, \tau^2_1, \rho\right)^\top$ and for AUSC using Pseudo-REML and Riley-REML when $\rho=0.3$ and $\tau^2_0=\tau^2_1=0.1$ (Tables \ref{tab:S3}-\ref{tab:S4});
\item Monte Carlo mean, Monte Carlo standard deviation, average of standard errors for parameters $\left(\alpha_0, \alpha_1, \gamma_0, \gamma_1, \tau^2_0, \tau^2_1, \rho\right)^\top$ and for AUSC using Pseudo-REML and Riley-REML when $\rho=0.9$ and $\tau^2_0=\tau^2_1=0.1$ (Tables \ref{tab:S5}-\ref{tab:S6});
\item Monte Carlo mean, Monte Carlo standard deviation, average of standard errors for the parameters $\left(\alpha_0, \alpha_1, \gamma_0, \gamma_1, \tau^2_0, \tau^2_1, \rho\right)^\top$ and for AUSC using Pseudo-ML and Riley-ML when $\rho =0.6$ and $\tau^2_0=\tau^2_1 = 0.1$ (Tables \ref{tab:S7}-\ref{tab:S8});

\item Empirical coverage of confidence intervals at nominal level 0.95 for parameters $\left(\alpha_0, \alpha_1, \gamma_0, \gamma_1\right)^\top$ and for AUSC using Pseudo-REML, Pseudo-ML, Riley-REML and Riley-ML when $\rho =0.6$ and $\tau^2_0=\tau^2_1 =1.0$ (Figures \ref{fig:S1}-\ref{fig:S2});
\item Empirical coverage of confidence intervals at nominal level 0.95 for parameters $\left(\alpha_0, \alpha_1, \gamma_0, \gamma_1\right)^\top$ and for AUSC using Pseudo-REML, Pseudo-ML, Riley-REML and Riley-ML when $\rho \in \{0.3, 0.9\}$ and $\tau^2_0=\tau^2_1 =0.1$ (Figures \ref{fig:S3}-\ref{fig:S6}).
\end{itemize}

\begin{sidewaystable}
	\centering
	\caption{Monte Carlo means (MCM), Monte Carlo standard deviations (MCSD), average of standard errors (ASE) for the parameters $\left(\alpha_0, \alpha_1, \gamma_0, \gamma_1, \tau^2_0, \tau^2_1, \rho\right)^\top$ and for AUSC, on the basis of 5,000 replicates of the simulation experiment with $\tau^2_{1} = \tau^2_{0} = 1, \rho = 0.6$ and increasing $K$. Maximum number of thresholds equal to 5.} 
	\label{tab:S1}
	\begingroup\scriptsize
	\begin{tabular}{lllrrrrrrrrrr}
		\toprule
		& & & & \multicolumn{3}{c}{$K = 10$} & \multicolumn{3}{c}{$K = 20$} & \multicolumn{3}{c}{$K = 50$}  \\
		\cmidrule(l){5-7} \cmidrule(l){8-10}  \cmidrule(l){11-13} 
		Thresholds& & & TRUE & MCM & MCSD & ASE & MCM & MCSD & ASE & MCM & MCSD & ASE \\
		\midrule
		\multirow{ 16 }{*}{ Full } & \multirow{ 8 }{*}{ Pseudo-REML } & $\alpha_1$ & 2.000 & 1.983 & 0.320 & 0.293 & 1.985 & 0.226 & 0.216 & 1.982 & 0.140 & 0.139 \\ 
		&  & $\alpha_0$ & 1.000 & 0.989 & 0.323 & 0.294 & 0.991 & 0.228 & 0.216 & 0.990 & 0.141 & 0.140 \\ 
		&  & $\gamma_1$ & $-$2.000 & $-$1.986 & 0.050 & 0.044 & $-$1.987 & 0.035 & 0.033 & $-$1.986 & 0.022 & 0.021 \\ 
		&  & $\gamma_0$ & 1.500 & 1.493 & 0.044 & 0.038 & 1.493 & 0.030 & 0.029 & 1.492 & 0.019 & 0.018 \\ 
		&  & $\tau^2_{1}$ & 1.000 & 0.990 & 0.464 & 0.373 & 0.990 & 0.319 & 0.289 & 0.984 & 0.200 & 0.191 \\ 
		&  & $\tau^2_{0}$ & 1.000 & 0.999 & 0.466 & 0.375 & 0.990 & 0.320 & 0.288 & 0.997 & 0.203 & 0.195 \\ 
		&  & $\rho$ & 0.600 & 0.566 & 0.244 & 0.177 & 0.579 & 0.159 & 0.137 & 0.585 & 0.097 & 0.091 \\ 
		&  & $\mathrm{AUSC}$ & 0.875 & 0.866 & 0.047 & 0.042 & 0.870 & 0.033 & 0.031 & 0.871 & 0.020 & 0.020 \\ 
		\cmidrule(l){2-13}
		& \multirow{ 8 }{*}{ Riley-REML } & $\alpha_1$ & 2.000 & 1.947 & 0.312 & 0.284 & 1.951 & 0.220 & 0.210 & 1.948 & 0.136 & 0.136 \\ 
		&  & $\alpha_0$ & 1.000 & 0.973 & 0.314 & 0.288 & 0.976 & 0.222 & 0.211 & 0.976 & 0.138 & 0.137 \\ 
		&  & $\gamma_1$ & $-$2.000 & $-$1.939 & 0.056 & 0.049 & $-$1.941 & 0.038 & 0.036 & $-$1.941 & 0.024 & 0.023 \\ 
		&  & $\gamma_0$ & 1.500 & 1.441 & 0.050 & 0.044 & 1.442 & 0.035 & 0.033 & 1.442 & 0.021 & 0.021 \\ 
		&  & $\tau^2_{1}$ & 1.000 & 0.954 & 0.454 & 0.366 & 0.953 & 0.311 & 0.282 & 0.949 & 0.196 & 0.187 \\ 
		&  & $\tau^2_{0}$ & 1.000 & 0.949 & 0.449 & 0.364 & 0.939 & 0.308 & 0.278 & 0.945 & 0.195 & 0.188 \\ 
		&  & $\rho$ & 0.600 & 0.578 & 0.247 & 0.180 & 0.591 & 0.161 & 0.138 & 0.597 & 0.098 & 0.092 \\ 
		&  & $\mathrm{AUSC}$ & 0.875 & 0.862 & 0.047 & 0.042 & 0.866 & 0.033 & 0.031 & 0.867 & 0.020 & 0.020 \\ 
		\midrule
		\multirow{ 16 }{*}{ Missing } & \multirow{ 8 }{*}{ Pseudo-REML } & $\alpha_1$ & 2.000 & 1.977 & 0.322 & 0.294 & 1.980 & 0.227 & 0.217 & 1.975 & 0.140 & 0.140 \\ 
		&  & $\alpha_0$ & 1.000 & 0.984 & 0.323 & 0.296 & 0.985 & 0.228 & 0.216 & 0.984 & 0.141 & 0.140 \\ 
		&  & $\gamma_1$ & $-$2.000 & $-$1.986 & 0.068 & 0.054 & $-$1.985 & 0.047 & 0.041 & $-$1.985 & 0.029 & 0.027 \\ 
		&  & $\gamma_0$ & 1.500 & 1.492 & 0.059 & 0.047 & 1.493 & 0.040 & 0.036 & 1.491 & 0.025 & 0.024 \\ 
		&  & $\tau^2_{1}$ & 1.000 & 0.974 & 0.465 & 0.377 & 0.975 & 0.321 & 0.291 & 0.969 & 0.201 & 0.192 \\ 
		&  & $\tau^2_{0}$ & 1.000 & 0.984 & 0.469 & 0.380 & 0.973 & 0.321 & 0.290 & 0.980 & 0.203 & 0.195 \\ 
		&  & $\rho$ & 0.600 & 0.571 & 0.253 & 0.183 & 0.583 & 0.163 & 0.140 & 0.588 & 0.099 & 0.093 \\ 
		&  & $\mathrm{AUSC}$ & 0.875 & 0.865 & 0.047 & 0.042 & 0.869 & 0.033 & 0.031 & 0.870 & 0.020 & 0.020 \\ 
		\cmidrule(l){2-13}
		& \multirow{ 8 }{*}{ Riley-REML } & $\alpha_1$ & 2.000 & 1.954 & 0.317 & 0.289 & 1.957 & 0.224 & 0.213 & 1.954 & 0.138 & 0.138 \\ 
		&  & $\alpha_0$ & 1.000 & 0.979 & 0.319 & 0.293 & 0.979 & 0.225 & 0.214 & 0.979 & 0.140 & 0.139 \\ 
		&  & $\gamma_1$ & $-$2.000 & $-$1.949 & 0.073 & 0.060 & $-$1.950 & 0.049 & 0.045 & $-$1.950 & 0.030 & 0.029 \\ 
		&  & $\gamma_0$ & 1.500 & 1.452 & 0.065 & 0.054 & 1.453 & 0.044 & 0.040 & 1.453 & 0.027 & 0.026 \\ 
		&  & $\tau^2_{1}$ & 1.000 & 0.951 & 0.459 & 0.373 & 0.952 & 0.317 & 0.287 & 0.946 & 0.198 & 0.190 \\ 
		&  & $\tau^2_{0}$ & 1.000 & 0.953 & 0.461 & 0.374 & 0.940 & 0.313 & 0.284 & 0.948 & 0.198 & 0.191 \\ 
		&  & $\rho$ & 0.600 & 0.580 & 0.255 & 0.185 & 0.593 & 0.165 & 0.141 & 0.597 & 0.100 & 0.094 \\ 
		&  & $\mathrm{AUSC}$ & 0.875 & 0.863 & 0.047 & 0.043 & 0.866 & 0.033 & 0.031 & 0.868 & 0.020 & 0.020 \\ 
		\bottomrule
	\end{tabular}
	\endgroup
\end{sidewaystable}

\begin{sidewaystable}
	\centering
	\caption{Monte Carlo means (MCM), Monte Carlo standard deviations (MCSD), average of standard errors (ASE) for the parameters $\left(\alpha_0, \alpha_1, \gamma_0, \gamma_1, \tau^2_0, \tau^2_1, \rho\right)^\top$ and for AUSC, on the basis of 5,000 replicates of the simulation experiment with $\tau^2_{1} = \tau^2_{0} = 1, \rho = 0.6$ and increasing $K$. Maximum number of thresholds equal to 15.} 
	\label{tab:S2}
	\begingroup\scriptsize
	\begin{tabular}{lllrrrrrrrrrr}
		\toprule
		& & & & \multicolumn{3}{c}{$K = 10$} & \multicolumn{3}{c}{$K = 20$} & \multicolumn{3}{c}{$K = 50$}  \\
		\cmidrule(l){5-7} \cmidrule(l){8-10}  \cmidrule(l){11-13} 
	Thresholds	& & & TRUE & MCM & MCSD & ASE & MCM & MCSD & ASE & MCM & MCSD & ASE \\
		\midrule
		\multirow{ 16 }{*}{Full} & \multirow{ 8 }{*}{ Pseudo-REML } & $\alpha_1$ & 2.000 & 1.988 & 0.316 & 0.294 & 1.986 & 0.223 & 0.216 & 1.985 & 0.141 & 0.140 \\ 
		&  & $\alpha_0$ & 1.000 & 0.996 & 0.320 & 0.294 & 0.994 & 0.227 & 0.216 & 0.994 & 0.142 & 0.141 \\ 
		&  & $\gamma_1$ & $-$2.000 & $-$1.991 & 0.051 & 0.045 & $-$1.990 & 0.035 & 0.034 & $-$1.990 & 0.022 & 0.022 \\ 
		&  & $\gamma_0$ & 1.500 & 1.496 & 0.045 & 0.039 & 1.494 & 0.031 & 0.030 & 1.495 & 0.020 & 0.019 \\ 
		&  & $\tau^2_{1}$ & 1.000 & 1.002 & 0.473 & 0.376 & 0.999 & 0.322 & 0.289 & 1.003 & 0.202 & 0.193 \\ 
		&  & $\tau^2_{0}$ & 1.000 & 1.011 & 0.482 & 0.378 & 1.001 & 0.328 & 0.291 & 1.015 & 0.210 & 0.197 \\ 
		&  & $\rho$ & 0.600 & 0.562 & 0.236 & 0.176 & 0.569 & 0.160 & 0.137 & 0.579 & 0.098 & 0.091 \\ 
		&  & $\mathrm{AUSC}$ & 0.875 & 0.867 & 0.046 & 0.042 & 0.870 & 0.032 & 0.030 & 0.872 & 0.020 & 0.020 \\ 
		\cmidrule(l){2-13}
		& \multirow{ 8 }{*}{ Riley-REML } & $\alpha_1$ & 2.000 & 1.823 & 0.280 & 0.261 & 1.826 & 0.197 & 0.192 & 1.830 & 0.125 & 0.125 \\ 
		&  & $\alpha_0$ & 1.000 & 0.941 & 0.291 & 0.270 & 0.941 & 0.208 & 0.198 & 0.942 & 0.130 & 0.129 \\ 
		&  & $\gamma_1$ & $-$2.000 & $-$1.784 & 0.088 & 0.076 & $-$1.788 & 0.060 & 0.055 & $-$1.791 & 0.036 & 0.036 \\ 
		&  & $\gamma_0$ & 1.500 & 1.289 & 0.079 & 0.069 & 1.290 & 0.057 & 0.051 & 1.294 & 0.038 & 0.033 \\ 
		&  & $\tau^2_{1}$ & 1.000 & 0.862 & 0.413 & 0.333 & 0.863 & 0.285 & 0.256 & 0.867 & 0.178 & 0.171 \\ 
		&  & $\tau^2_{0}$ & 1.000 & 0.857 & 0.418 & 0.331 & 0.848 & 0.284 & 0.254 & 0.860 & 0.181 & 0.171 \\ 
		&  & $\rho$ & 0.600 & 0.582 & 0.243 & 0.180 & 0.589 & 0.163 & 0.139 & 0.598 & 0.098 & 0.092 \\ 
		&  & $\mathrm{AUSC}$ & 0.875 & 0.849 & 0.045 & 0.043 & 0.852 & 0.032 & 0.031 & 0.854 & 0.020 & 0.020 \\ 
		\midrule
		\multirow{ 16 }{*}{Missing} & \multirow{ 8 }{*}{ Pseudo-REML } & $\alpha_1$ & 2.000 & 1.984 & 0.316 & 0.295 & 1.982 & 0.223 & 0.216 & 1.982 & 0.141 & 0.140 \\ 
		&  & $\alpha_0$ & 1.000 & 0.993 & 0.320 & 0.295 & 0.991 & 0.227 & 0.216 & 0.990 & 0.142 & 0.141 \\ 
		&  & $\gamma_1$ & $-$2.000 & $-$1.989 & 0.062 & 0.051 & $-$1.989 & 0.043 & 0.039 & $-$1.990 & 0.027 & 0.026 \\ 
		&  & $\gamma_0$ & 1.500 & 1.495 & 0.055 & 0.044 & 1.494 & 0.038 & 0.034 & 1.495 & 0.024 & 0.023 \\ 
		&  & $\tau^2_{1}$ & 1.000 & 0.993 & 0.472 & 0.379 & 0.988 & 0.323 & 0.290 & 0.993 & 0.202 & 0.194 \\ 
		&  & $\tau^2_{0}$ & 1.000 & 1.001 & 0.483 & 0.381 & 0.990 & 0.328 & 0.292 & 1.002 & 0.210 & 0.197 \\ 
		&  & $\rho$ & 0.600 & 0.565 & 0.242 & 0.178 & 0.572 & 0.163 & 0.139 & 0.581 & 0.099 & 0.092 \\ 
		&  & $\mathrm{AUSC}$ & 0.875 & 0.867 & 0.046 & 0.042 & 0.869 & 0.032 & 0.031 & 0.871 & 0.020 & 0.020 \\ 
		\cmidrule(l){2-13}
		& \multirow{ 8 }{*}{ Riley-REML } & $\alpha_1$ & 2.000 & 1.891 & 0.296 & 0.276 & 1.892 & 0.208 & 0.203 & 1.894 & 0.132 & 0.132 \\ 
		&  & $\alpha_0$ & 1.000 & 0.970 & 0.305 & 0.283 & 0.967 & 0.216 & 0.207 & 0.968 & 0.136 & 0.135 \\ 
		&  & $\gamma_1$ & $-$2.000 & $-$1.859 & 0.084 & 0.071 & $-$1.862 & 0.058 & 0.052 & $-$1.864 & 0.034 & 0.034 \\ 
		&  & $\gamma_0$ & 1.500 & 1.362 & 0.076 & 0.064 & 1.362 & 0.055 & 0.048 & 1.365 & 0.034 & 0.032 \\ 
		&  & $\tau^2_{1}$ & 1.000 & 0.912 & 0.439 & 0.357 & 0.909 & 0.303 & 0.272 & 0.914 & 0.189 & 0.182 \\ 
		&  & $\tau^2_{0}$ & 1.000 & 0.910 & 0.446 & 0.356 & 0.896 & 0.302 & 0.271 & 0.908 & 0.192 & 0.181 \\ 
		&  & $\rho$ & 0.600 & 0.581 & 0.247 & 0.182 & 0.589 & 0.165 & 0.141 & 0.597 & 0.099 & 0.093 \\ 
		&  & $\mathrm{AUSC}$ & 0.875 & 0.857 & 0.046 & 0.043 & 0.860 & 0.032 & 0.031 & 0.862 & 0.020 & 0.020 \\ 
		\bottomrule
	\end{tabular}
	\endgroup
\end{sidewaystable}

\begin{sidewaystable}[h]
	\centering
\caption{Monte Carlo mean (MCM), Monte Carlo standard deviation (MCSD), average of standard errors (ASE) for the parameters $\left(\alpha_0, \alpha_1, \gamma_0, \gamma_1, \tau^2_0, \tau^2_1, \rho\right)^\top$ and for AUSC, on the basis of 5,000 replicates of the simulation experiment with $\tau^2_{1} = \tau^2_{0} = 0.1, \rho = 0.3$ and increasing $K$. Maximum number of thresholds equal to 5.}
\label{tab:S3}
\begingroup\scriptsize
\begin{tabular}{lllrrrrrrrrrr}
\toprule
& & & & \multicolumn{3}{c}{$K = 10$} & \multicolumn{3}{c}{$K = 20$} & \multicolumn{3}{c}{$K = 50$}  \\
\cmidrule(l){5-7} \cmidrule(l){8-10}  \cmidrule(l){11-13}
Thresholds& & & TRUE & MCM & MCSD & ASE & MCM & MCSD & ASE & MCM & MCSD & ASE \\
\midrule
\multirow{ 16 }{*}{Full } & \multirow{ 8 }{*}{Pseudo-REML} & $\alpha_1$ & 2.000 & 1.983 & 0.118 & 0.107 & 1.986 & 0.083 & 0.079 & 1.985 & 0.051 & 0.051 \\
&  & $\alpha_0$ & 1.000 & 0.992 & 0.113 & 0.104 & 0.994 & 0.078 & 0.076 & 0.993 & 0.049 & 0.049 \\
&  & $\gamma_1$ & $-$2.000 & $-$1.989 & 0.048 & 0.043 & $-$1.988 & 0.034 & 0.032 & $-$1.988 & 0.021 & 0.021 \\
&  & $\gamma_0$ & 1.500 & 1.494 & 0.043 & 0.038 & 1.493 & 0.030 & 0.028 & 1.493 & 0.019 & 0.018 \\
&  & $\tau^2_{1}$ & 0.100 & 0.111 & 0.058 & 0.046 & 0.112 & 0.040 & 0.035 & 0.111 & 0.025 & 0.023 \\
&  & $\tau^2_{0}$ & 0.100 & 0.117 & 0.061 & 0.048 & 0.115 & 0.041 & 0.037 & 0.114 & 0.026 & 0.024 \\
&  & $\rho$ & 0.300 & 0.258 & 0.356 & 0.259 & 0.254 & 0.241 & 0.204 & 0.258 & 0.145 & 0.138 \\
&  & $\mathrm{AUSC}$ & 0.875 & 0.872 & 0.014 & 0.013 & 0.873 & 0.010 & 0.009 & 0.873 & 0.006 & 0.006 \\
\cmidrule(l){2-13}
& \multirow{ 8 }{*}{ Riley-REML } & $\alpha_1$ & 2.000 & 1.950 & 0.116 & 0.104 & 1.954 & 0.081 & 0.077 & 1.953 & 0.050 & 0.050 \\
&  & $\alpha_0$ & 1.000 & 0.977 & 0.110 & 0.101 & 0.979 & 0.075 & 0.074 & 0.978 & 0.048 & 0.048 \\
&  & $\gamma_1$ & $-$2.000 & $-$1.949 & 0.050 & 0.045 & $-$1.949 & 0.035 & 0.034 & $-$1.949 & 0.022 & 0.022 \\
&  & $\gamma_0$ & 1.500 & 1.448 & 0.047 & 0.041 & 1.448 & 0.033 & 0.031 & 1.448 & 0.020 & 0.020 \\
&  & $\tau^2_{1}$ & 0.100 & 0.095 & 0.055 & 0.043 & 0.096 & 0.037 & 0.033 & 0.096 & 0.023 & 0.022 \\
&  & $\tau^2_{0}$ & 0.100 & 0.097 & 0.056 & 0.045 & 0.096 & 0.038 & 0.033 & 0.095 & 0.024 & 0.022 \\
&  & $\rho$ & 0.300 & 0.303 & 0.425 & 0.329 & 0.295 & 0.276 & 0.228 & 0.299 & 0.163 & 0.153 \\
&  & $\mathrm{AUSC}$ & 0.875 & 0.868 & 0.014 & 0.013 & 0.869 & 0.010 & 0.009 & 0.869 & 0.006 & 0.006 \\
\midrule
\multirow{ 16 }{*}{ Missing  } & \multirow{ 8 }{*}{ Pseudo-REML } & $\alpha_1$ & 2.000 & 1.982 & 0.124 & 0.112 & 1.984 & 0.085 & 0.081 & 1.983 & 0.053 & 0.053 \\
&  & $\alpha_0$ & 1.000 & 0.991 & 0.116 & 0.107 & 0.992 & 0.078 & 0.077 & 0.991 & 0.050 & 0.050 \\
&  & $\gamma_1$ & $-$2.000 & $-$1.988 & 0.063 & 0.054 & $-$1.987 & 0.044 & 0.040 & $-$1.986 & 0.027 & 0.027 \\
&  & $\gamma_0$ & 1.500 & 1.494 & 0.058 & 0.047 & 1.492 & 0.039 & 0.035 & 1.492 & 0.025 & 0.023 \\
&  & $\tau^2_{1}$ & 0.100 & 0.107 & 0.061 & 0.048 & 0.107 & 0.041 & 0.036 & 0.107 & 0.026 & 0.024 \\
&  & $\tau^2_{0}$ & 0.100 & 0.111 & 0.063 & 0.050 & 0.110 & 0.042 & 0.037 & 0.109 & 0.026 & 0.025 \\
&  & $\rho$ & 0.300 & 0.267 & 0.415 & 0.384 & 0.262 & 0.269 & 0.222 & 0.268 & 0.161 & 0.150 \\
&  & $\mathrm{AUSC}$ & 0.875 & 0.872 & 0.015 & 0.013 & 0.872 & 0.010 & 0.010 & 0.872 & 0.006 & 0.006 \\
\cmidrule(l){2-13}
& \multirow{ 8 }{*}{ Riley-REML } & $\alpha_1$ & 2.000 & 1.960 & 0.122 & 0.121 & 1.963 & 0.084 & 0.080 & 1.962 & 0.052 & 0.052 \\
&  & $\alpha_0$ & 1.000 & 0.985 & 0.114 & 0.109 & 0.986 & 0.077 & 0.076 & 0.985 & 0.049 & 0.049 \\
&  & $\gamma_1$ & $-$2.000 & $-$1.958 & 0.064 & 0.059 & $-$1.957 & 0.046 & 0.042 & $-$1.957 & 0.028 & 0.027 \\
&  & $\gamma_0$ & 1.500 & 1.458 & 0.060 & 0.053 & 1.457 & 0.041 & 0.037 & 1.457 & 0.025 & 0.024 \\
&  & $\tau^2_{1}$ & 0.100 & 0.096 & 0.059 & 0.054 & 0.097 & 0.040 & 0.035 & 0.096 & 0.025 & 0.023 \\
&  & $\tau^2_{0}$ & 0.100 & 0.098 & 0.059 & 0.055 & 0.096 & 0.040 & 0.035 & 0.096 & 0.025 & 0.023 \\
&  & $\rho$ & 0.300 & 0.300 & 0.471 & 0.615 & 0.295 & 0.301 & 0.246 & 0.299 & 0.176 & 0.163 \\
&  & $\mathrm{AUSC}$ & 0.875 & 0.869 & 0.015 & 0.014 & 0.870 & 0.010 & 0.010 & 0.870 & 0.006 & 0.006 \\
\bottomrule
\end{tabular}
\endgroup
\end{sidewaystable}

\begin{sidewaystable}[htbp]
	\centering
	\caption{Monte Carlo mean (MCM), Monte Carlo standard deviation (MCSD), average of standard errors (ASE) for the parameters $\left(\alpha_0, \alpha_1, \gamma_0, \gamma_1, \tau^2_0, \tau^2_1, \rho\right)^\top$ and for AUSC, on the basis of 5,000 replicates of the simulation experiment with $\tau^2_{1} = \tau^2_{0} = 0.1, \rho = 0.3$ and increasing $K$. Maximum number of thresholds equal to 15.} 
	\label{tab:S4}
	\begingroup\scriptsize
	\begin{tabular}{lllrrrrrrrrrr}
		\toprule
		Thresholds& & & & \multicolumn{3}{c}{$K = 10$} & \multicolumn{3}{c}{$K = 20$} & \multicolumn{3}{c}{$K = 50$}  \\
		\cmidrule(l){5-7} \cmidrule(l){8-10}  \cmidrule(l){11-13} 
		& & & TRUE & MCM & MCSD & ASE & MCM & MCSD & ASE & MCM & MCSD & ASE \\
		\midrule
		\multirow{ 16 }{*}{ Full  } & \multirow{ 8 }{*}{ Pseudo-REML } & $\alpha_1$ & 2.000 & 1.993 & 0.118 & 0.108 & 1.991 & 0.084 & 0.080 & 1.990 & 0.052 & 0.051 \\ 
		&  & $\alpha_0$ & 1.000 & 1.000 & 0.114 & 0.105 & 0.999 & 0.081 & 0.077 & 0.998 & 0.050 & 0.050 \\ 
		&  & $\gamma_1$ & $-$2.000 & $-$1.993 & 0.050 & 0.044 & $-$1.991 & 0.035 & 0.033 & $-$1.992 & 0.022 & 0.021 \\ 
		&  & $\gamma_0$ & 1.500 & 1.497 & 0.044 & 0.040 & 1.496 & 0.032 & 0.029 & 1.496 & 0.020 & 0.019 \\ 
		&  & $\tau^2_{1}$ & 0.100 & 0.121 & 0.061 & 0.048 & 0.121 & 0.042 & 0.037 & 0.120 & 0.026 & 0.024 \\ 
		&  & $\tau^2_{0}$ & 0.100 & 0.127 & 0.064 & 0.050 & 0.125 & 0.044 & 0.038 & 0.125 & 0.027 & 0.026 \\ 
		&  & $\rho$ & 0.300 & 0.230 & 0.329 & 0.244 & 0.236 & 0.227 & 0.194 & 0.239 & 0.140 & 0.131 \\ 
		&  & $\mathrm{AUSC}$ & 0.875 & 0.873 & 0.014 & 0.013 & 0.874 & 0.010 & 0.010 & 0.874 & 0.006 & 0.006 \\ 
		\cmidrule(l){2-13}
		& \multirow{ 8 }{*}{ Riley-REML } & $\alpha_1$ & 2.000 & 1.837 & 0.112 & 0.104 & 1.837 & 0.079 & 0.075 & 1.838 & 0.049 & 0.049 \\ 
		&  & $\alpha_0$ & 1.000 & 0.939 & 0.103 & 0.095 & 0.940 & 0.073 & 0.070 & 0.939 & 0.046 & 0.045 \\ 
		&  & $\gamma_1$ & $-$2.000 & $-$1.808 & 0.072 & 0.064 & $-$1.809 & 0.051 & 0.047 & $-$1.811 & 0.031 & 0.030 \\ 
		&  & $\gamma_0$ & 1.500 & 1.302 & 0.069 & 0.060 & 1.305 & 0.048 & 0.044 & 1.305 & 0.030 & 0.029 \\ 
		&  & $\tau^2_{1}$ & 0.100 & 0.086 & 0.050 & 0.041 & 0.087 & 0.034 & 0.030 & 0.086 & 0.021 & 0.020 \\ 
		&  & $\tau^2_{0}$ & 0.100 & 0.087 & 0.051 & 0.041 & 0.087 & 0.034 & 0.030 & 0.086 & 0.022 & 0.020 \\ 
		&  & $\rho$ & 0.300 & 0.300 & 0.434 & 0.362 & 0.304 & 0.277 & 0.232 & 0.304 & 0.168 & 0.155 \\ 
		&  & $\mathrm{AUSC}$ & 0.875 & 0.855 & 0.014 & 0.013 & 0.855 & 0.010 & 0.010 & 0.855 & 0.006 & 0.006 \\ 
		\midrule
		\multirow{ 16 }{*}{ Missing  } & \multirow{ 8 }{*}{ Pseudo-REML } & $\alpha_1$ & 2.000 & 1.993 & 0.120 & 0.110 & 1.990 & 0.085 & 0.081 & 1.988 & 0.053 & 0.052 \\ 
		&  & $\alpha_0$ & 1.000 & 0.998 & 0.115 & 0.106 & 0.998 & 0.081 & 0.077 & 0.997 & 0.051 & 0.050 \\ 
		&  & $\gamma_1$ & $-$2.000 & $-$1.994 & 0.061 & 0.050 & $-$1.991 & 0.042 & 0.038 & $-$1.991 & 0.026 & 0.026 \\ 
		&  & $\gamma_0$ & 1.500 & 1.497 & 0.054 & 0.044 & 1.496 & 0.038 & 0.034 & 1.495 & 0.024 & 0.023 \\ 
		&  & $\tau^2_{1}$ & 0.100 & 0.117 & 0.062 & 0.048 & 0.117 & 0.042 & 0.037 & 0.115 & 0.026 & 0.024 \\ 
		&  & $\tau^2_{0}$ & 0.100 & 0.121 & 0.064 & 0.050 & 0.119 & 0.044 & 0.038 & 0.120 & 0.027 & 0.026 \\ 
		&  & $\rho$ & 0.300 & 0.239 & 0.352 & 0.259 & 0.245 & 0.240 & 0.204 & 0.247 & 0.147 & 0.138 \\ 
		&  & $\mathrm{AUSC}$ & 0.875 & 0.873 & 0.014 & 0.013 & 0.873 & 0.010 & 0.010 & 0.873 & 0.006 & 0.006 \\ 
		\cmidrule(l){2-13}
		& \multirow{ 8 }{*}{ Riley-REML } & $\alpha_1$ & 2.000 & 1.904 & 0.116 & 0.151 & 1.902 & 0.081 & 0.078 & 1.902 & 0.051 & 0.050 \\ 
		&  & $\alpha_0$ & 1.000 & 0.971 & 0.109 & 0.102 & 0.971 & 0.077 & 0.073 & 0.970 & 0.048 & 0.047 \\ 
		&  & $\gamma_1$ & $-$2.000 & $-$1.877 & 0.072 & 0.069 & $-$1.876 & 0.049 & 0.046 & $-$1.877 & 0.031 & 0.030 \\ 
		&  & $\gamma_0$ & 1.500 & 1.372 & 0.068 & 0.060 & 1.374 & 0.048 & 0.043 & 1.373 & 0.029 & 0.028 \\ 
		&  & $\tau^2_{1}$ & 0.100 & 0.093 & 0.054 & 0.053 & 0.093 & 0.037 & 0.033 & 0.091 & 0.023 & 0.022 \\ 
		&  & $\tau^2_{0}$ & 0.100 & 0.094 & 0.055 & 0.067 & 0.093 & 0.037 & 0.033 & 0.092 & 0.023 & 0.022 \\ 
		&  & $\rho$ & 0.300 & 0.297 & 0.448 & 0.699 & 0.300 & 0.285 & 0.238 & 0.300 & 0.171 & 0.158 \\ 
		&  & $\mathrm{AUSC}$ & 0.875 & 0.863 & 0.014 & 0.016 & 0.864 & 0.010 & 0.010 & 0.864 & 0.006 & 0.006 \\ 
		\bottomrule
	\end{tabular}
	\endgroup
\end{sidewaystable}

\begin{sidewaystable}[h]
	\centering
	\caption{Monte Carlo mean (MCM), Monte Carlo standard deviation (MCSD), average of standard errors (ASE) for the parameters $\left(\alpha_0, \alpha_1, \gamma_0, \gamma_1, \tau^2_0, \tau^2_1, \rho\right)^\top$ and for AUSC, on the basis of 5,000 replicates of the simulation experiment with $\tau^2_{1} = \tau^2_{0} = 0.1, \rho = 0.9$ and increasing $K$. Maximum number of thresholds equal to 5.} 
	\label{tab:S5}
	\begingroup\scriptsize
	\begin{tabular}{lllrrrrrrrrrr}
		\toprule
		& & & & \multicolumn{3}{c}{$K = 10$} & \multicolumn{3}{c}{$K = 20$} & \multicolumn{3}{c}{$K = 50$}  \\
		\cmidrule(l){5-7} \cmidrule(l){8-10}  \cmidrule(l){11-13} 
		Thresholds& & & TRUE & MCM & MCSD & ASE & MCM & MCSD & ASE & MCM & MCSD & ASE \\
		\midrule
		\multirow{ 16 }{*}{ Full} & \multirow{ 8 }{*}{ Pseudo-REML } & $\alpha_1$ & 2.000 & 1.990 & 0.117 & 0.107 & 1.985 & 0.082 & 0.078 & 1.985 & 0.051 & 0.050 \\ 
		&  & $\alpha_0$ & 1.000 & 0.997 & 0.110 & 0.104 & 0.994 & 0.078 & 0.075 & 0.994 & 0.049 & 0.048 \\ 
		&  & $\gamma_1$ & $-$2.000 & $-$1.989 & 0.049 & 0.043 & $-$1.988 & 0.034 & 0.032 & $-$1.988 & 0.021 & 0.021 \\ 
		&  & $\gamma_0$ & 1.500 & 1.493 & 0.042 & 0.038 & 1.494 & 0.029 & 0.028 & 1.493 & 0.019 & 0.018 \\ 
		&  & $\tau^2_{1}$ & 0.100 & 0.109 & 0.058 & 0.048 & 0.110 & 0.039 & 0.035 & 0.109 & 0.024 & 0.023 \\ 
		&  & $\tau^2_{0}$ & 0.100 & 0.114 & 0.059 & 0.050 & 0.113 & 0.040 & 0.035 & 0.112 & 0.025 & 0.024 \\ 
		&  & $\rho$ & 0.900 & 0.780 & 0.193 & 0.148 & 0.792 & 0.116 & 0.099 & 0.796 & 0.069 & 0.064 \\ 
		&  & $\mathrm{AUSC}$ & 0.875 & 0.873 & 0.016 & 0.015 & 0.873 & 0.012 & 0.011 & 0.873 & 0.007 & 0.007 \\ 
		\cmidrule(l){2-13}
		& \multirow{ 8 }{*}{ Riley-REML } & $\alpha_1$ & 2.000 & 1.957 & 0.114 & 0.121 & 1.953 & 0.081 & 0.078 & 1.953 & 0.050 & 0.049 \\ 
		&  & $\alpha_0$ & 1.000 & 0.983 & 0.107 & 0.110 & 0.979 & 0.076 & 0.074 & 0.980 & 0.048 & 0.047 \\ 
		&  & $\gamma_1$ & $-$2.000 & $-$1.950 & 0.052 & 0.048 & $-$1.949 & 0.036 & 0.034 & $-$1.949 & 0.022 & 0.022 \\ 
		&  & $\gamma_0$ & 1.500 & 1.448 & 0.046 & 0.041 & 1.449 & 0.032 & 0.031 & 1.448 & 0.020 & 0.020 \\ 
		&  & $\tau^2_{1}$ & 0.100 & 0.095 & 0.054 & 0.086 & 0.097 & 0.037 & 0.041 & 0.095 & 0.023 & 0.022 \\ 
		&  & $\tau^2_{0}$ & 0.100 & 0.096 & 0.055 & 0.089 & 0.096 & 0.037 & 0.044 & 0.095 & 0.023 & 0.022 \\ 
		&  & $\rho$ & 0.900 & 0.895 & 0.172 & 0.429 & 0.907 & 0.095 & 0.129 & 0.907 & 0.058 & 0.054 \\ 
		&  & $\mathrm{AUSC}$ & 0.875 & 0.869 & 0.016 & 0.017 & 0.869 & 0.012 & 0.011 & 0.869 & 0.007 & 0.007 \\ 
		\midrule
		\multirow{ 16 }{*}{ Missing  } & \multirow{ 8 }{*}{ Pseudo-REML } & $\alpha_1$ & 2.000 & 1.989 & 0.121 & 0.111 & 1.984 & 0.085 & 0.081 & 1.983 & 0.053 & 0.052 \\ 
		&  & $\alpha_0$ & 1.000 & 0.996 & 0.112 & 0.108 & 0.993 & 0.079 & 0.077 & 0.994 & 0.050 & 0.049 \\ 
		&  & $\gamma_1$ & $-$2.000 & $-$1.987 & 0.064 & 0.056 & $-$1.987 & 0.044 & 0.041 & $-$1.986 & 0.027 & 0.026 \\ 
		&  & $\gamma_0$ & 1.500 & 1.493 & 0.057 & 0.049 & 1.492 & 0.038 & 0.036 & 1.492 & 0.024 & 0.023 \\ 
		&  & $\tau^2_{1}$ & 0.100 & 0.106 & 0.061 & 0.052 & 0.106 & 0.040 & 0.038 & 0.105 & 0.025 & 0.023 \\ 
		&  & $\tau^2_{0}$ & 0.100 & 0.109 & 0.062 & 0.056 & 0.109 & 0.041 & 0.038 & 0.108 & 0.025 & 0.024 \\ 
		&  & $\rho$ & 0.900 & 0.807 & 0.224 & 0.227 & 0.818 & 0.128 & 0.118 & 0.819 & 0.074 & 0.067 \\ 
		&  & $\mathrm{AUSC}$ & 0.875 & 0.873 & 0.016 & 0.015 & 0.872 & 0.012 & 0.011 & 0.873 & 0.007 & 0.007 \\ 
		\cmidrule(l){2-13}
		& \multirow{ 8 }{*}{ Riley-REML } & $\alpha_1$ & 2.000 & 1.968 & 0.119 & 0.128 & 1.963 & 0.084 & 0.085 & 1.963 & 0.052 & 0.051 \\ 
		&  & $\alpha_0$ & 1.000 & 0.991 & 0.110 & 0.123 & 0.987 & 0.077 & 0.083 & 0.987 & 0.049 & 0.048 \\ 
		&  & $\gamma_1$ & $-$2.000 & $-$1.959 & 0.065 & 0.068 & $-$1.958 & 0.046 & 0.046 & $-$1.958 & 0.028 & 0.027 \\ 
		&  & $\gamma_0$ & 1.500 & 1.460 & 0.058 & 0.059 & 1.459 & 0.039 & 0.040 & 1.459 & 0.025 & 0.024 \\ 
		&  & $\tau^2_{1}$ & 0.100 & 0.097 & 0.058 & 0.078 & 0.097 & 0.039 & 0.065 & 0.096 & 0.024 & 0.023 \\ 
		&  & $\tau^2_{0}$ & 0.100 & 0.098 & 0.058 & 0.096 & 0.097 & 0.039 & 0.060 & 0.096 & 0.024 & 0.023 \\ 
		&  & $\rho$ & 0.900 & 0.884 & 0.216 & 0.788 & 0.904 & 0.110 & 0.290 & 0.907 & 0.065 & 0.063 \\ 
		&  & $\mathrm{AUSC}$ & 0.875 & 0.870 & 0.016 & 0.017 & 0.870 & 0.012 & 0.012 & 0.870 & 0.007 & 0.007 \\ 
		\bottomrule
	\end{tabular}
	\endgroup
\end{sidewaystable}

\begin{sidewaystable}[h]
	\centering
	\caption{Monte Carlo mean (MCM), Monte Carlo standard deviation (MCSD), average of standard errors (ASE) for the parameters $\left(\alpha_0, \alpha_1, \gamma_0, \gamma_1, \tau^2_0, \tau^2_1, \rho\right)^\top$ and for AUSC, on the basis of 5,000 replicates of the simulation experiment with $\tau^2_{1} = \tau^2_{0} = 0.1, \rho = 0.9$ and increasing $K$. Maximum number of thresholds equal to 15.} 
	\label{tab:S6}
	\begingroup\scriptsize
	\begin{tabular}{lllrrrrrrrrrr}
		\toprule
		& & & & \multicolumn{3}{c}{$K = 10$} & \multicolumn{3}{c}{$K = 20$} & \multicolumn{3}{c}{$K = 50$}  \\
		\cmidrule(l){5-7} \cmidrule(l){8-10}  \cmidrule(l){11-13} 
	Thresholds	& & & TRUE & MCM & MCSD & ASE & MCM & MCSD & ASE & MCM & MCSD & ASE \\
		\midrule
		\multirow{ 16 }{*}{ Full  } & \multirow{ 8 }{*}{ Pseudo-REML } & $\alpha_1$ & 2.000 & 1.993 & 0.117 & 0.108 & 1.990 & 0.082 & 0.079 & 1.990 & 0.051 & 0.051 \\ 
		&  & $\alpha_0$ & 1.000 & 0.998 & 0.110 & 0.104 & 0.997 & 0.080 & 0.077 & 0.998 & 0.051 & 0.049 \\ 
		&  & $\gamma_1$ & $-$2.000 & $-$1.993 & 0.050 & 0.045 & $-$1.992 & 0.034 & 0.033 & $-$1.991 & 0.021 & 0.021 \\ 
		&  & $\gamma_0$ & 1.500 & 1.496 & 0.044 & 0.039 & 1.496 & 0.031 & 0.029 & 1.496 & 0.020 & 0.019 \\ 
		&  & $\tau^2_{1}$ & 0.100 & 0.121 & 0.060 & 0.048 & 0.120 & 0.041 & 0.036 & 0.119 & 0.025 & 0.024 \\ 
		&  & $\tau^2_{0}$ & 0.100 & 0.123 & 0.062 & 0.049 & 0.123 & 0.042 & 0.038 & 0.123 & 0.027 & 0.025 \\ 
		&  & $\rho$ & 0.900 & 0.715 & 0.199 & 0.141 & 0.728 & 0.128 & 0.108 & 0.732 & 0.078 & 0.072 \\ 
		&  & $\mathrm{AUSC}$ & 0.875 & 0.873 & 0.016 & 0.015 & 0.873 & 0.012 & 0.011 & 0.874 & 0.007 & 0.007 \\ 
		\cmidrule(l){2-13}
		& \multirow{ 8 }{*}{ Riley-REML } & $\alpha_1$ & 2.000 & 1.839 & 0.111 & 0.112 & 1.837 & 0.077 & 0.080 & 1.838 & 0.049 & 0.048 \\ 
		&  & $\alpha_0$ & 1.000 & 0.940 & 0.100 & 0.105 & 0.939 & 0.072 & 0.073 & 0.940 & 0.046 & 0.044 \\ 
		&  & $\gamma_1$ & $-$2.000 & $-$1.809 & 0.072 & 0.064 & $-$1.810 & 0.050 & 0.047 & $-$1.810 & 0.030 & 0.030 \\ 
		&  & $\gamma_0$ & 1.500 & 1.303 & 0.070 & 0.059 & 1.304 & 0.050 & 0.044 & 1.306 & 0.035 & 0.029 \\ 
		&  & $\tau^2_{1}$ & 0.100 & 0.088 & 0.049 & 0.066 & 0.087 & 0.034 & 0.039 & 0.087 & 0.021 & 0.020 \\ 
		&  & $\tau^2_{0}$ & 0.100 & 0.087 & 0.050 & 0.062 & 0.087 & 0.034 & 0.038 & 0.087 & 0.022 & 0.020 \\ 
		&  & $\rho$ & 0.900 & 0.907 & 0.159 & 0.433 & 0.918 & 0.091 & 0.265 & 0.919 & 0.057 & 0.054 \\ 
		&  & $\mathrm{AUSC}$ & 0.875 & 0.855 & 0.017 & 0.017 & 0.855 & 0.012 & 0.011 & 0.855 & 0.008 & 0.007 \\ 
		\midrule
		\multirow{ 16 }{*}{ Missing } & \multirow{ 8 }{*}{ Pseudo-REML } & $\alpha_1$ & 2.000 & 1.992 & 0.119 & 0.110 & 1.989 & 0.083 & 0.080 & 1.988 & 0.052 & 0.052 \\ 
		&  & $\alpha_0$ & 1.000 & 0.997 & 0.110 & 0.104 & 0.997 & 0.080 & 0.077 & 0.997 & 0.051 & 0.049 \\ 
		&  & $\gamma_1$ & $-$2.000 & $-$1.993 & 0.061 & 0.050 & $-$1.991 & 0.041 & 0.038 & $-$1.991 & 0.026 & 0.025 \\ 
		&  & $\gamma_0$ & 1.500 & 1.496 & 0.055 & 0.044 & 1.495 & 0.037 & 0.034 & 1.495 & 0.023 & 0.022 \\ 
		&  & $\tau^2_{1}$ & 0.100 & 0.117 & 0.060 & 0.049 & 0.115 & 0.041 & 0.036 & 0.115 & 0.025 & 0.024 \\ 
		&  & $\tau^2_{0}$ & 0.100 & 0.118 & 0.062 & 0.050 & 0.118 & 0.042 & 0.037 & 0.118 & 0.027 & 0.025 \\ 
		&  & $\rho$ & 0.900 & 0.742 & 0.207 & 0.149 & 0.754 & 0.130 & 0.108 & 0.757 & 0.078 & 0.071 \\ 
		&  & $\mathrm{AUSC}$ & 0.875 & 0.873 & 0.016 & 0.015 & 0.873 & 0.012 & 0.011 & 0.873 & 0.007 & 0.007 \\ 
		\cmidrule(l){2-13}
		& \multirow{ 8 }{*}{ Riley-REML } & $\alpha_1$ & 2.000 & 1.905 & 0.114 & 0.124 & 1.903 & 0.080 & 0.078 & 1.902 & 0.050 & 0.050 \\ 
		&  & $\alpha_0$ & 1.000 & 0.971 & 0.105 & 0.166 & 0.970 & 0.075 & 0.073 & 0.971 & 0.048 & 0.047 \\ 
		&  & $\gamma_1$ & $-$2.000 & $-$1.878 & 0.073 & 0.072 & $-$1.878 & 0.049 & 0.046 & $-$1.877 & 0.030 & 0.030 \\ 
		&  & $\gamma_0$ & 1.500 & 1.373 & 0.069 & 0.070 & 1.374 & 0.049 & 0.043 & 1.375 & 0.029 & 0.028 \\ 
		&  & $\tau^2_{1}$ & 0.100 & 0.094 & 0.054 & 0.067 & 0.093 & 0.037 & 0.039 & 0.093 & 0.023 & 0.022 \\ 
		&  & $\tau^2_{0}$ & 0.100 & 0.093 & 0.054 & 0.074 & 0.093 & 0.037 & 0.041 & 0.092 & 0.023 & 0.022 \\ 
		&  & $\rho$ & 0.900 & 0.893 & 0.182 & 1.791 & 0.909 & 0.100 & 0.157 & 0.910 & 0.061 & 0.059 \\ 
		&  & $\mathrm{AUSC}$ & 0.875 & 0.863 & 0.016 & 0.020 & 0.863 & 0.012 & 0.011 & 0.864 & 0.007 & 0.007 \\ 
		\bottomrule
	\end{tabular}
	\endgroup
\end{sidewaystable}


\begin{sidewaystable}[h]
	\centering
	\caption{Monte Carlo mean (MCM), Monte Carlo standard deviation (MCSD), average of standard errors (ASE) for the parameters $\left(\alpha_0, \alpha_1, \gamma_0, \gamma_1, \tau^2_0, \tau^2_1, \rho\right)^\top$ and for AUSC, on the basis of 5,000 replicates of the simulation experiment with $\tau^2_{1} = \tau^2_{0} = 0.1, \rho = 0.6$ and increasing $K$. Maximum number of thresholds equal to 5.} 
	\label{tab:S7}
	\begingroup\scriptsize
	\begin{tabular}{lllrrrrrrrrrr}
		\toprule
		& & & & \multicolumn{3}{c}{$K = 10$} & \multicolumn{3}{c}{$K = 20$} & \multicolumn{3}{c}{$K = 50$}  \\
		\cmidrule(l){5-7} \cmidrule(l){8-10}  \cmidrule(l){11-13} 
	Thresholds	& & & TRUE & MCM & MCSD & ASE & MCM & MCSD & ASE & MCM & MCSD & ASE \\
		\midrule
		\multirow{ 16 }{*}{ Full } & \multirow{ 8 }{*}{ Pseudo-ML } & $\alpha_1$ & 2.000 & 1.985 & 0.115 & 0.107 & 1.987 & 0.079 & 0.079 & 1.986 & 0.052 & 0.051 \\ 
		&  & $\alpha_0$ & 1.000 & 0.996 & 0.111 & 0.103 & 0.995 & 0.076 & 0.076 & 0.993 & 0.049 & 0.049 \\ 
		&  & $\gamma_1$ & $-$2.000 & $-$1.989 & 0.048 & 0.043 & $-$1.988 & 0.033 & 0.032 & $-$1.988 & 0.021 & 0.021 \\ 
		&  & $\gamma_0$ & 1.500 & 1.494 & 0.043 & 0.038 & 1.493 & 0.030 & 0.028 & 1.493 & 0.019 & 0.018 \\ 
		&  & $\tau^2_{1}$ & 0.100 & 0.099 & 0.052 & 0.041 & 0.105 & 0.037 & 0.033 & 0.108 & 0.023 & 0.023 \\ 
		&  & $\tau^2_{0}$ & 0.100 & 0.101 & 0.052 & 0.042 & 0.108 & 0.039 & 0.035 & 0.111 & 0.025 & 0.024 \\ 
		&  & $\rho$ & 0.600 & 0.521 & 0.301 & 0.219 & 0.524 & 0.195 & 0.167 & 0.523 & 0.122 & 0.111 \\ 
		&  & $\mathrm{AUSC}$ & 0.875 & 0.872 & 0.015 & 0.014 & 0.873 & 0.010 & 0.010 & 0.873 & 0.007 & 0.007 \\ 
		\cmidrule(l){2-13}
		& \multirow{ 8 }{*}{ Riley-ML } & $\alpha_1$ & 2.000 & 1.952 & 0.114 & 0.105 & 1.954 & 0.078 & 0.077 & 1.954 & 0.051 & 0.049 \\ 
		&  & $\alpha_0$ & 1.000 & 0.980 & 0.108 & 0.101 & 0.980 & 0.074 & 0.073 & 0.978 & 0.048 & 0.047 \\ 
		&  & $\gamma_1$ & $-$2.000 & $-$1.949 & 0.052 & 0.046 & $-$1.949 & 0.035 & 0.034 & $-$1.949 & 0.022 & 0.022 \\ 
		&  & $\gamma_0$ & 1.500 & 1.447 & 0.047 & 0.041 & 1.448 & 0.032 & 0.031 & 1.448 & 0.020 & 0.020 \\ 
		&  & $\tau^2_{1}$ & 0.100 & 0.085 & 0.048 & 0.039 & 0.090 & 0.035 & 0.031 & 0.093 & 0.022 & 0.021 \\ 
		&  & $\tau^2_{0}$ & 0.100 & 0.084 & 0.049 & 0.040 & 0.090 & 0.036 & 0.032 & 0.093 & 0.023 & 0.022 \\ 
		&  & $\rho$ & 0.600 & 0.614 & 0.344 & 0.264 & 0.614 & 0.214 & 0.178 & 0.604 & 0.129 & 0.117 \\ 
		&  & $\mathrm{AUSC}$ & 0.875 & 0.868 & 0.015 & 0.014 & 0.869 & 0.010 & 0.010 & 0.869 & 0.007 & 0.007 \\ 
		\midrule
		\multirow{ 16 }{*}{ Missing } & \multirow{ 8 }{*}{ Pseudo-ML } & $\alpha_1$ & 2.000 & 1.983 & 0.121 & 0.111 & 1.985 & 0.083 & 0.081 & 1.984 & 0.053 & 0.052 \\ 
		&  & $\alpha_0$ & 1.000 & 0.994 & 0.113 & 0.105 & 0.993 & 0.078 & 0.077 & 0.992 & 0.050 & 0.050 \\ 
		&  & $\gamma_1$ & $-$2.000 & $-$1.987 & 0.065 & 0.053 & $-$1.987 & 0.044 & 0.040 & $-$1.986 & 0.027 & 0.026 \\ 
		&  & $\gamma_0$ & 1.500 & 1.493 & 0.058 & 0.046 & 1.492 & 0.039 & 0.035 & 1.492 & 0.024 & 0.023 \\ 
		&  & $\tau^2_{1}$ & 0.100 & 0.094 & 0.055 & 0.043 & 0.100 & 0.039 & 0.034 & 0.103 & 0.024 & 0.023 \\ 
		&  & $\tau^2_{0}$ & 0.100 & 0.095 & 0.054 & 0.043 & 0.102 & 0.040 & 0.035 & 0.106 & 0.025 & 0.024 \\ 
		&  & $\rho$ & 0.600 & 0.549 & 0.356 & 0.252 & 0.548 & 0.218 & 0.181 & 0.541 & 0.133 & 0.120 \\ 
		&  & $\mathrm{AUSC}$ & 0.875 & 0.872 & 0.015 & 0.014 & 0.873 & 0.011 & 0.010 & 0.873 & 0.007 & 0.007 \\ 
		\cmidrule(l){2-13}
		& \multirow{ 8 }{*}{ Riley-ML } & $\alpha_1$ & 2.000 & 1.962 & 0.120 & 0.110 & 1.964 & 0.082 & 0.080 & 1.963 & 0.052 & 0.051 \\ 
		&  & $\alpha_0$ & 1.000 & 0.987 & 0.112 & 0.105 & 0.987 & 0.076 & 0.076 & 0.985 & 0.049 & 0.049 \\ 
		&  & $\gamma_1$ & $-$2.000 & $-$1.958 & 0.067 & 0.056 & $-$1.958 & 0.045 & 0.042 & $-$1.957 & 0.028 & 0.027 \\ 
		&  & $\gamma_0$ & 1.500 & 1.458 & 0.060 & 0.049 & 1.458 & 0.041 & 0.037 & 1.458 & 0.025 & 0.024 \\ 
		&  & $\tau^2_{1}$ & 0.100 & 0.085 & 0.052 & 0.042 & 0.090 & 0.037 & 0.033 & 0.093 & 0.024 & 0.022 \\ 
		&  & $\tau^2_{0}$ & 0.100 & 0.083 & 0.051 & 0.042 & 0.090 & 0.038 & 0.033 & 0.093 & 0.024 & 0.023 \\ 
		&  & $\rho$ & 0.600 & 0.613 & 0.389 & 0.334 & 0.620 & 0.238 & 0.195 & 0.606 & 0.141 & 0.126 \\ 
		&  & $\mathrm{AUSC}$ & 0.875 & 0.870 & 0.015 & 0.014 & 0.870 & 0.011 & 0.010 & 0.870 & 0.007 & 0.007 \\ 
		\bottomrule
	\end{tabular}
	\endgroup
\end{sidewaystable}

\begin{sidewaystable}[h]
	\centering
	\caption{Monte Carlo mean (MCM), Monte Carlo standard deviation (MCSD), average of standard errors (ASE) for the parameters $\left(\alpha_0, \alpha_1, \gamma_0, \gamma_1, \tau^2_0, \tau^2_1, \rho\right)^\top$ and for AUSC, on the basis of 5,000 replicates of the simulation experiment with $\tau^2_{1} = \tau^2_{0} = 0.1, \rho = 0.6$ and increasing $K$. Maximum number of thresholds equal to 15.} 
	\label{tab:S8}
	\begingroup\scriptsize
	\begin{tabular}{lllrrrrrrrrrr}
		\toprule
		& & & & \multicolumn{3}{c}{$K = 10$} & \multicolumn{3}{c}{$K = 20$} & \multicolumn{3}{c}{$K = 50$}  \\
		\cmidrule(l){5-7} \cmidrule(l){8-10}  \cmidrule(l){11-13} 
	Thresholds	& & & TRUE & MCM & MCSD & ASE & MCM & MCSD & ASE & MCM & MCSD & ASE \\
		\midrule
		\multirow{ 16 }{*}{ Full } & \multirow{ 8 }{*}{ Pseudo-ML } & $\alpha_1$ & 2.000 & 1.989 & 0.118 & 0.108 & 1.991 & 0.083 & 0.079 & 1.989 & 0.052 & 0.051 \\ 
		&  & $\alpha_0$ & 1.000 & 0.997 & 0.113 & 0.105 & 0.997 & 0.078 & 0.077 & 0.998 & 0.050 & 0.050 \\ 
		&  & $\gamma_1$ & $-$2.000 & $-$1.993 & 0.049 & 0.044 & $-$1.991 & 0.034 & 0.033 & $-$1.991 & 0.022 & 0.021 \\ 
		&  & $\gamma_0$ & 1.500 & 1.497 & 0.044 & 0.040 & 1.496 & 0.031 & 0.029 & 1.496 & 0.019 & 0.019 \\ 
		&  & $\tau^2_{1}$ & 0.100 & 0.108 & 0.054 & 0.043 & 0.113 & 0.039 & 0.035 & 0.118 & 0.025 & 0.024 \\ 
		&  & $\tau^2_{0}$ & 0.100 & 0.112 & 0.058 & 0.044 & 0.118 & 0.042 & 0.036 & 0.121 & 0.026 & 0.025 \\ 
		&  & $\rho$ & 0.600 & 0.466 & 0.288 & 0.209 & 0.473 & 0.193 & 0.164 & 0.481 & 0.117 & 0.110 \\ 
		&  & $\mathrm{AUSC}$ & 0.875 & 0.873 & 0.015 & 0.014 & 0.873 & 0.011 & 0.010 & 0.874 & 0.007 & 0.007 \\ 
		\cmidrule(l){2-13}
		& \multirow{ 8 }{*}{ Riley-ML } & $\alpha_1$ & 2.000 & 1.833 & 0.112 & 0.104 & 1.837 & 0.079 & 0.075 & 1.837 & 0.049 & 0.049 \\ 
		&  & $\alpha_0$ & 1.000 & 0.937 & 0.103 & 0.096 & 0.938 & 0.071 & 0.069 & 0.939 & 0.046 & 0.045 \\ 
		&  & $\gamma_1$ & $-$2.000 & $-$1.807 & 0.071 & 0.064 & $-$1.808 & 0.049 & 0.047 & $-$1.810 & 0.031 & 0.030 \\ 
		&  & $\gamma_0$ & 1.500 & 1.301 & 0.069 & 0.059 & 1.305 & 0.048 & 0.044 & 1.306 & 0.031 & 0.029 \\ 
		&  & $\tau^2_{1}$ & 0.100 & 0.076 & 0.045 & 0.035 & 0.081 & 0.033 & 0.028 & 0.084 & 0.021 & 0.020 \\ 
		&  & $\tau^2_{0}$ & 0.100 & 0.076 & 0.046 & 0.036 & 0.081 & 0.033 & 0.029 & 0.084 & 0.021 & 0.020 \\ 
		&  & $\rho$ & 0.600 & 0.609 & 0.364 & 0.285 & 0.615 & 0.224 & 0.182 & 0.613 & 0.127 & 0.118 \\ 
		&  & $\mathrm{AUSC}$ & 0.875 & 0.854 & 0.016 & 0.014 & 0.855 & 0.011 & 0.010 & 0.855 & 0.007 & 0.007 \\ 
		\midrule
		\multirow{ 16 }{*}{ Missing } & \multirow{ 8 }{*}{ Pseudo-ML } & $\alpha_1$ & 2.000 & 1.989 & 0.120 & 0.109 & 1.989 & 0.084 & 0.080 & 1.988 & 0.053 & 0.052 \\ 
		&  & $\alpha_0$ & 1.000 & 0.996 & 0.113 & 0.105 & 0.996 & 0.078 & 0.077 & 0.996 & 0.050 & 0.050 \\ 
		&  & $\gamma_1$ & $-$2.000 & $-$1.992 & 0.059 & 0.050 & $-$1.990 & 0.042 & 0.038 & $-$1.991 & 0.026 & 0.025 \\ 
		&  & $\gamma_0$ & 1.500 & 1.496 & 0.053 & 0.044 & 1.495 & 0.037 & 0.034 & 1.496 & 0.023 & 0.023 \\ 
		&  & $\tau^2_{1}$ & 0.100 & 0.103 & 0.054 & 0.043 & 0.109 & 0.040 & 0.035 & 0.113 & 0.025 & 0.024 \\ 
		&  & $\tau^2_{0}$ & 0.100 & 0.107 & 0.058 & 0.045 & 0.112 & 0.041 & 0.036 & 0.116 & 0.026 & 0.025 \\ 
		&  & $\rho$ & 0.600 & 0.488 & 0.310 & 0.221 & 0.492 & 0.205 & 0.171 & 0.499 & 0.121 & 0.114 \\ 
		&  & $\mathrm{AUSC}$ & 0.875 & 0.873 & 0.015 & 0.014 & 0.873 & 0.011 & 0.010 & 0.873 & 0.007 & 0.007 \\ 
		\cmidrule(l){2-13}
		& \multirow{ 8 }{*}{ Riley-ML } & $\alpha_1$ & 2.000 & 1.900 & 0.116 & 0.107 & 1.902 & 0.081 & 0.077 & 1.902 & 0.051 & 0.050 \\ 
		&  & $\alpha_0$ & 1.000 & 0.970 & 0.108 & 0.101 & 0.970 & 0.074 & 0.073 & 0.970 & 0.048 & 0.047 \\ 
		&  & $\gamma_1$ & $-$2.000 & $-$1.876 & 0.072 & 0.062 & $-$1.876 & 0.049 & 0.046 & $-$1.877 & 0.031 & 0.030 \\ 
		&  & $\gamma_0$ & 1.500 & 1.372 & 0.066 & 0.057 & 1.373 & 0.047 & 0.043 & 1.375 & 0.029 & 0.028 \\ 
		&  & $\tau^2_{1}$ & 0.100 & 0.080 & 0.048 & 0.039 & 0.086 & 0.035 & 0.031 & 0.090 & 0.022 & 0.021 \\ 
		&  & $\tau^2_{0}$ & 0.100 & 0.081 & 0.050 & 0.040 & 0.086 & 0.035 & 0.031 & 0.090 & 0.022 & 0.021 \\ 
		&  & $\rho$ & 0.600 & 0.604 & 0.378 & 0.291 & 0.610 & 0.233 & 0.190 & 0.606 & 0.130 & 0.122 \\ 
		&  & $\mathrm{AUSC}$ & 0.875 & 0.863 & 0.015 & 0.014 & 0.863 & 0.011 & 0.010 & 0.864 & 0.007 & 0.007 \\ 
		\bottomrule
	\end{tabular}
	\endgroup
\end{sidewaystable}




\clearpage
\section{Additional example: The glycated haemoglobin A1c} 
\begin{figure}[h]
	\begin{center}
		\includegraphics[width=3.5in]{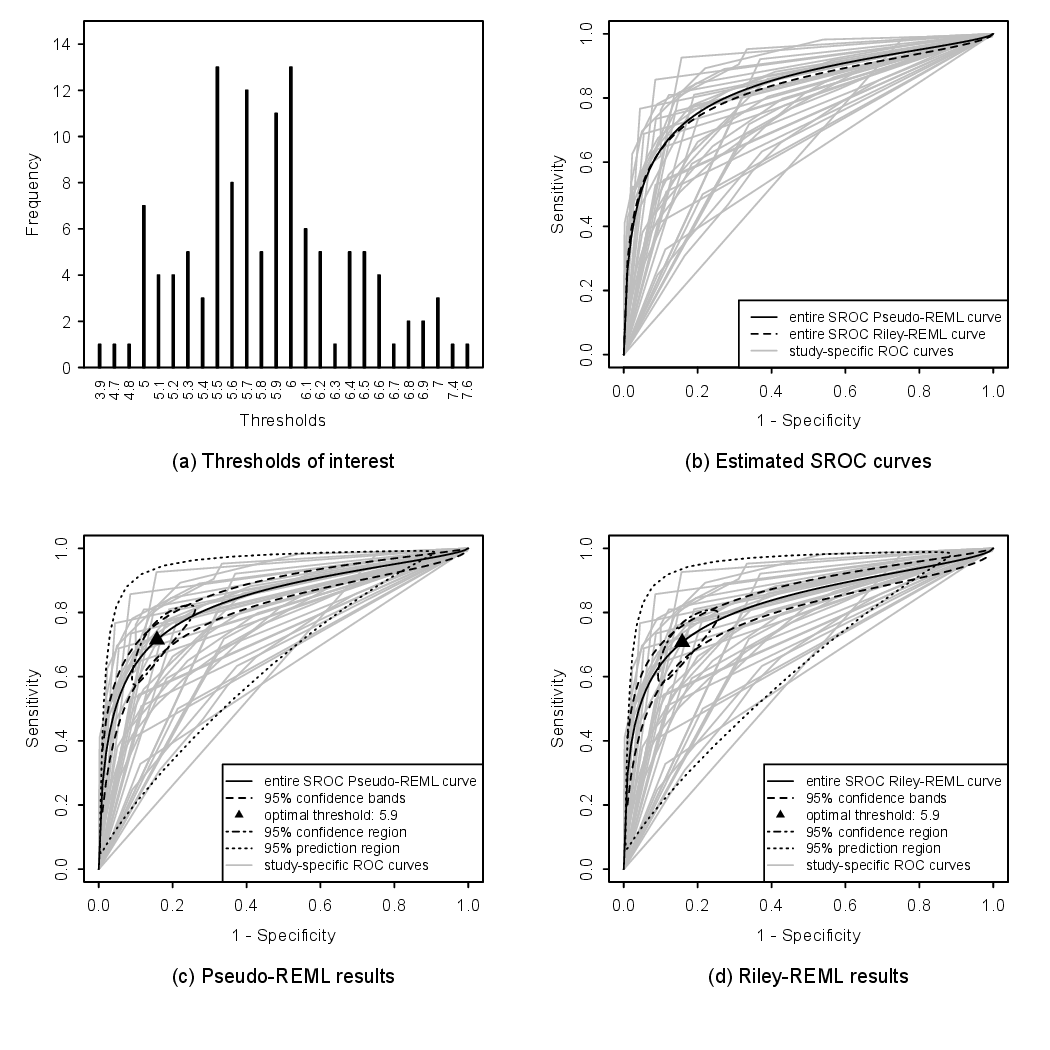}
	\end{center}
	\caption{Glycated haemoglobin A1c data analysis. Panel (a): frequency of the thresholds. Panel (b): study-specific and estimated SROC curves. Panels (c)--(d): SROC curve with 95\% confidence bands; optimal sensitivity and 1-specificity with 95\% confidence region; prediction region. Methods: Riley-REML and Pseudo-REML.}
	\label{fig:HbA1c}
\end{figure}

The measure of glycated haemoglobin A1c as a test to identify type 2 diabetes mellitus has been considered an alternative to the reference standard represented by glucose measurement through fasting plasma glucose or oral glucose tolerance test. Recently, Hoyer et al. \cite{hoyer2017meta} performed a meta-analysis of 38 studies with 124 pairs of sensitivity and specificity from 26 different thresholds ranging from 3.9\% to 7.6\% obtained from previous analyses in Bennet et al. \cite{bennett2007hba1c} and Kodama et al. \cite{kodama2013use}. The frequency of thresholds in each study is shown in Figure \ref{fig:HbA1c}(a).
\begin{table}[h]
	\begin{center}
		\caption{Glycated haemoglobin A1c data. Results for the parameters in model (2.2) in the main paper and for AUSC from using Pseudo-REML and Riley-REML.}
		\label{tab:Hb1Ac}
		\begin{tabular}{l r r r | r r r}
			\toprule
			& \multicolumn{3}{c|}{Pseudo-REML} & \multicolumn{3}{c}{Riley-REML} \\
			\midrule
			& Estimate & Std. error & $p$-value & Estimate & Std. error & $p$-value \\
			\midrule
			$\alpha_1$ & 18.605 & 2.053 & $<$ 0.001 & 15.960 & 1.738 & $<$ 0.001 \\
			$\alpha_0$ & $-$24.899 & 1.137 & $<$ 0.001 & $-$23.441 & 1.572 & $<$ 0.001 \\
			$\gamma_1$ & $-$2.998 & 0.350 & $<$ 0.001 & $-$2.556 & 0.301 & $<$ 0.001 \\
			$\gamma_0$ & 4.504 & 0.196 & $<$ 0.001 & 4.257 & 0.274 & $<$ 0.001  \\
			$\tau_1^2$ & 2.454 & 0.924 & -- & 2.092 & 0.780 & -- \\
			$\tau_0^2$ & 2.576 & 1.041 & -- & 2.261 & 0.978 & -- \\
			$\rho$ & $-$0.882 & 0.052 & -- & $-$0.862 & 0.060 & -- \\
			\midrule[1.5pt]
			& Estimate & Std. error & 95\% interval& Estimate & Std. error & 95\% interval \\
			\midrule
			AUSC & 0.837 & 0.112 & (0.506, 0.963) & 0.827 & 0.159 & (0.351, 0.977) \\
			\bottomrule
		\end{tabular}
	\end{center}
\end{table}

Table \ref{tab:Hb1Ac} reports estimates, sandwich estimates of standard errors and p-values for the test of significance of the parameters of interest using Riley et al. \cite{riley2014meta2} approach and the pseudo-likelihood approach, both under REML estimation. All the regression coefficients are significantly different from zero at 5\% level, under both the approaches. Large values of variance components indicate the presence of heterogeneity between studies for all the thresholds. The negative between-study correlation indicate that studies with higher sensitivity are likely to have lower specificity.

The SROC curves from Pseudo-REML and Riley-REML reported in Figure \ref{fig:HbA1c}(b) show a similar behaviour and, as a consequence, values of AUSC are comparable, with only a slightly larger and less variable result from Pseudo-REML. The optimal threshold based on the Youden index is 5.9\% for both the approaches. Summary sensitivities and specificities are equal to 71.5\% and 84.3\%, respectively, for Pseudo-REML, and to 70.7\% and 84.2\%, respectively, for Riley-REML. The values are reported into the ROC space in Figure \ref{fig:HbA1c}(c)-(d) for Pseudo-REML and Riley-REML, respectively, together with the associated 95\% confidence regions and 95\% prediction regions. In the same graph, the 95\% confidence bands of the SROC curves are superimposed. Result from both the approaches are comparable to those in Hoyer et al. \cite{hoyer2017meta}, who consider a bivariate time-to-event models for interval-censored data.

\end{document}